# Interaction of electron beams with optical nanostructures and metamaterials: From coherent photon sources towards shaping the wave function

Nahid Talebi

Stuttgart Center for Electron Microscopy, Max Planck Institute for Solid State Research, Heisenbergstr. 1, 70569 Stuttgart

Abstract- Investigating the interaction of electron beams with materials and light has been a field of research since more than a century. The field was advanced theoretically by the raise of quantum mechanics and technically by the introduction of electron microscopes and accelerators. It is possible nowadays to uncover a multitude of information from electron-induced excitations in matter by means of advanced techniques like holography, tomography, and most recently photon-induced near-field electron microscopy. The question is whether the interaction can be controlled in an even more efficient way in order to unravel important questions like modal decomposition of the electron-induced polarization, by performing experiments with better spatial, temporal, and energy resolutions. This review discusses recent advances in controlling the electron and light interactions at the nanoscale. Theoretical and numerical aspects of the interaction of electrons with nanostructures and metamaterials will be discussed, with the aim to understand mechanisms of radiation in interaction of electrons with even more sophisticated structures. Based on these mechanisms of radiation, state-of-the art and novel electron-driven few-photon sources will be discussed. Applications of such sources to gain an understanding of quantum optical effects and also to perform spectral interferometry with electron microscopes will be covered. In an inverse approach, as in the case of the inverse Smith–Purcell effect, laser-induced excitations of nanostructures can cause the electron beams traveling in the near-field of such structures to get accelerated, provided a synchronization criterion is satisfied. This effect is the basis for linear dielectric and metallic electron accelerators. Moreover, acceleration goes along with bunching of the electrons. When single electrons are considered, an efficient design of nanostructures can lead to the shaping of the electron wave function travelling adjacent to them, for example to form attosecond electron pulses or chiral electron wave functions.

Keywords- Electron beam, Wave function, Radiation, Interference, Few-photon source, Acceleration, Beam shaping

**Outline-**

1. Introduction

About one century after the introduction of the Schrödinger equation and the invention of the transmission electron microscope by Ernst Ruska, electron microscopy has entered nowadays the era of ultrahigh sub-Ångstrom spatial resolution [1]. However, microscopy of materials at high spatial resolution is not the only possible application of electron microscopes; indeed electron spectroscopy is almost playing the same important role for material characterization. Nanoscience is the science of understanding the dynamical physical and chemical processes which happen at the nanoscale, within the time–energy and momentum–space phase spaces. For this reason, electron probes offering enough brightness, emittance, and spatiotemporal coherence length are crucial for our understanding and manipulation of the nanoworld.

Interestingly, similar to light, swift electrons can also undergo inelastic interaction with single electrons and by collective electron excitations within the sample, like plasmon and photon polaritons, as a result of which they will lose energy [2]. Within the classical formalism, the electron energy-loss (EEL) spectrum is theoretically rationalized by a simple but intuitive interpretation, which has a direct correspondence with the basic principles of quantum mechanics, demanding that all inelastic signal is collected [3]. In the non-recoil approximation and assuming that the electron is travelling parallel to the $z$-axis, the EEL spectrum is provided as $\Gamma^{\text{EELS}}(\omega) = (e/\pi\hbar\omega) \times \text{Re} \int dz\, E_z^{\text{sca}}(R_e, z; \omega) \exp(-i z \omega/V) = (e/\pi\hbar\omega) \text{Re}\, \tilde{E}_z^{\text{sca}}(R_e; k_z = \omega/V; \omega)$, where $e$ is the elementary charge, $\hbar$ is the reduced Planck constant, $\omega$ is the angular frequency of the light, $R_e = (x_e, y_e)$ is the electron impact parameter, $\vec{V} = V\hat{z}$ is the velocity of the electron, $k_z$ is the $z$-component of the wave vector, and $\tilde{E}_z^{\text{sca}}$ is the Fourier-transformed $z$-component of the electric field. Moreover, by measuring the amount of energy loss of the electrons using an EEL detector, we have a direct access to the photonic local density of states projected along the electron trajectory [4]. Considering this formalism, many aspects of the optical near-field distribution of nanostructures, like dark versus bright modes [5-8], localized surface plasmons [9-12], void plasmons [13-16], and wedge plasmons [17-19] are understood by EEL spectroscopy (EELS). The specific selection of the wave vector ($k_z = \omega/V$) by swift electrons takes into account the energy–momentum conservation criterion which is crucial for mapping the whole local density of mesoscopic structures such as gratings [20] and conical tapers [21, 22].

An electron interacting with materials and photonic modes of the sample causes electromagnetic radiation to the far-field, which can be analyzed using cathodoluminescence (CL) spectroscopy [23-25]. CL is complementary to EELS, in the sense that only the radiative modes are detected by CL. However, optically dark modes such as bulk plasmons still contribute to the electron-induced radiation [26], because the emitter (here the electron) traverses the material. Moreover, whenever an electron beam crosses an interface between a metal and a dielectric, transition radiation (TR) as well as plasmon radiation contribute to the far-field, which are both mutually coherent with the evanescent near-field of the electron [2].

Indeed many aspects of the interaction of the electrons with nanostructures are well understood and investigated in the literature, covering a broad range of material science, plasmonics, and acceleration physics. This review provides a rather general overview of the whole field, with emphasis on provoking further research on the interaction of electron beams with carefully engineered structures, such as gratings and metamaterials [27, 28], either to create novel few photon sources or to shape the electron beam, but also to provide new characterization methodologies by combing these two seemingly distinct aspects of the field. We, however, intentionally exclude the aspects of photon-induced near-field electron microscopy [29] which has recently attracted a lot of attention, as this research field has been already reviewed in recent articles [30]. We start with a brief introduction to the plasmonics and metamaterials, which is then followed by detailed theoretical explanations of electron-photon interactions.

## 2. Mechanisms of radiation of the electron interacting with nanostructures and metamaterials

*2.1. Plasmonics and metametrials*

Bulk plasmons are quanta of collective oscillations of free electrons in the form of electron-density waves, due to the long-range Coulomb interaction between valence electrons in metals [31, 32]. Additionally, existence of boundaries such as surfaces of a thin film, can lead to the propagation of the plasmons and its coupling with photons at the surface, hence results in the formation of surface plasmon polaritons (SPPs). In other words, SPP unlike Bulk plasmon can carry the electromagnetic energy. Interestingly, SPPs have been theoretically predicted by Ritchie [33] and shortly after experimentally investigated by Powell and Swan [34, 35] using electron probes in an electron microscope. From those early days, electron microscopes have been considered as an important tool for characterization of collective excitations in metals.

Considering the electrodynamics, an interface between two isotropic nonmagnetic materials with dielectric functions $\varepsilon_1(\omega)$ and $\varepsilon_2(\omega)$ can only support a bound optical mode if $\varepsilon_1(\omega) < -\varepsilon_2(\omega)$ [31]. Evidently, this criterion can be met by inclusion of materials like noble metals and their alloys positioned adjacent to a dielectric. Additionally, electronic transitions due to the excitation of phonons and excitons, as well as Cooper pairs in superconductors can also lead to the satisfaction the above-mentioned criterion [36]. Moreover, the propagation constant of the SPP at the interface of two isotropic materials is also given by $\beta = k_0 \sqrt{\varepsilon_1(\omega)\varepsilon_2(\omega)/(\varepsilon_1(\omega) + \varepsilon_2(\omega))}$, where $k_0 = \omega/c$ is the free space wavenumber, $\omega$ is the angular frequency and $c$ is the speed of light in vacuum. Interesting aspects of SPP include transmission and confinement of the electromagnetic energy beyond the diffraction limit [37] and field enhancement up to several orders of magnitude [38], which pave the way towards nanocicuitry [32] and nonlinear optics [39]. Due to the capability of SPPs for ultrahigh confinement of optical energies, plasmonic resonators can be scaled down to few nanometers [40]. Plasmons in nanoresonators are usually called localized plasmons, and have several applications in nanosensing [41], Nanoantennas [42], and enhancing the Purcell factor [43].

The field of plasmonics involves engineering of optical waves using natural materials and their given dielectric function. Additionally, engineering of novel materials to conduct the flow of optical waves in a more controlled way have led to the emergence of spoof plasmons [44, 45] and metamaterials [27, 46-48]. Whereas the former is achieved by corrugated surfaces, metamaterials are precisely designed nanostructures in two and three dimensional arrays to engineer not only the effective dielectric function, but also the effective permeability. The general effort in the design of metamaterials is to create artificial materials with concomitant negative permittivity and permeability, due to the application of such a material in controlling the phase of the propagating light at surfaces [27] or at three dimensions [49].

## 2.2. Basic principles for electron – photon interactions

Interaction of electrons and optical waves in infinite free space is only restricted to *elastic* scattering, a prominent example of which is the Kapitza–Dirac effect (KDE), as pointed out by Kapitza and Dirac in 1933 [50]. In the KDE, an electron beam interacts with a standing wave light pattern in free space. As a result of such an interaction the electron beam is elastically diffracted, mainly because of a two-photon absorption and emission process [51]. Similar elastic scattering effects for ionic beams have been utilized in matter wave experiments, using standing-wave optical patterns as gratings to efficiently split the matter waves into the required beam paths [52, 53]. Moreover, travelling-wave-assisted beam splitters have also been proposed and investigated theoretically, demonstrating that the effect is not necessarily restricted to standing wave patterns in free space [54]. However, because of the small values of the Thomson scattering cross section, quite high intensities of the laser beam are required to achieve intense diffracted matter waves. For this reason the experimental verification of the KDE effect had to wait for the development of the laser [55]. Figure 1 demonstrates the diffraction of a Gaussian electron wave function from two propagating Gaussian optical waves at the wavelength of 30 nm. The processes were simulated with the self-consisted numerical toolbox explained in Section 6. The electron has a velocity of $0.04c$, longitudinal and transverse widths of 5 nm, and is travelling through the interference point of the optical waves, as shown in figure 1(a). The amplitude of the electric field is $E_0 = 1.5 \times 10^{12} V/m$. Evolution of the electron wave function in spatial representation is shown in figure 1(b), which clearly reveals the diffraction of the wave function into three peaks by light.

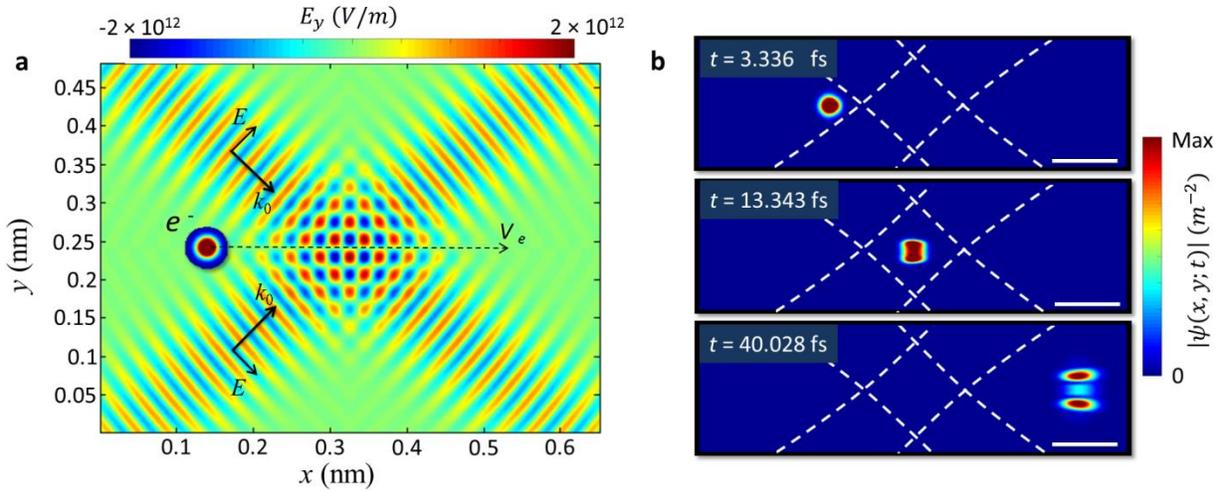

**Figure 1.** A Gaussian electron wave packet interacting with two propagating Gaussian optical waves, demonstrating the KDE. (a) Spatial profile of the *y*-component of the electric field at a given time. The wavelength of the light wave is $\lambda_0 = 30 \, \text{nm}$, and the polarization state is depicted in the frame. The electron wave packet traverses the optical waves at their intersection at a velocity of $0.04c$. (b) The magnitude of the electron wave function at times depicted at each frame. The scale bar is 100 nm.

By contrast, for *inelastic* electron–photon interaction to occur, one needs a mediator to satisfy both the energy and momentum conservation criteria. For the case of the emission of a single photon by an electron, these criteria are given by

$$\begin{cases} E_f = E_i + \hbar\omega \\ \vec{p}_f = \vec{p}_i + \hbar\vec{k} \end{cases} \quad (1)$$

assuming that other selection rules, as for the angular momentum, are satisfied. In eq. (1) $E_\alpha$ is the electron energy, $\alpha \in (i, f)$ denotes the initial and final states of the electron before and after the interaction, $\omega$ is the angular frequency of the emitted photon, $\vec{k}$ is the photon wave vector, and $\vec{p}_i$ and $\vec{p}_f$ are the linear momenta of the electron before and after the interaction, respectively. Based upon the relativistic energies $E_\alpha = \sqrt{(m_0^2 c^4 + p_\alpha^2 c^2)}$ and momenta $\vec{p}_\alpha = \gamma_\alpha m_0 \vec{v}_\alpha$, where $\gamma_\alpha$ is the Lorentz factor and $\vec{v}_\alpha$ is the electron velocity, the critical angle is inferred as

$$\cos(\theta) = \frac{(\hbar k)^2 - (\hbar k_0)^2}{2 p_i \hbar k} \pm \frac{c k_0}{v_i k} \quad (2)$$

where $\cos(\theta) = \vec{p}_i \cdot \vec{k} / p_i k$ and $k_0$ is the free-space wavenumber of the photon. It is evident that in the free-space ($k = k_0$), eq. (2) will simplify to $\cos(\theta) = \pm c/v_i$, which cannot lead to real values for $\theta$. However, there exist several ways to satisfy eq. (2) [56, 57], either by manipulating the momentum of the photon or the electron, or their energies, in either longitudinal or transverse directions. Among these possibilities, slow-wave mediators incorporating optical devices for manipulating the momentum of the photons allow for most effective designs, and are discussed here. For devices which use Larmor-based mechanisms of radiation and which incorporate quasi-free electrons, it is referred to [58-61] for free electron lasers and [62, 63] for cyclotron radiation.

In the classical limit where $\hbar \to 0$ and also at low photon energies where $\hbar\omega \ll p_i c$ the second term in eq. (2) is dominant. In order to have $c/v_i \leq k/k_0$, one can use a matterial with refractive index $n(\omega) \geq c/v_i$. This is the condition for Cherenkov radiation (CR) [64], and the underlying mechanism will be discussed in more detail in Section 2.3. Another method is to make use of the nearfield of materials which provide evanescent modes with wavenumbers large enough to satisfy eq. (2). An example of such a topology is an electron traveling adjacent and parallel to a single interface [65]. Moreover, when an electron crosses an interface, this leads to TR [2, 66]. The origin of this radiation is understood by the coupling of the moving electron with its image charge inside the material, which together form a transient dipole. Examples of TR from metamaterials will be shown in Section 2.4. The Smith–Purcell effect occurs when an electron travels near a grating, as will be discussed in section 2.5.

*2.3. Cherenkov radiation*

CR takes place as a result of the interaction of a moving charge with a bulk material with refractive index higher than $c/v$, where $v$ is the electron velocity. Despite being similar to the acoustic wave created by a uniformly moving source at a velocity larger than the sound velocity, it was only in the 1930's that an equivalent electromagnetic wave was experimentally detected [67-69]. Evidently this is due to the fact that accelerators were needed to create sufficiently fast moving charges.

Several publications were devoted to the classical [64, 70] and quantum-mechanical [71-74] theoretical basis of CR and also its applications [75] in isotropic and homogeneous metals and dielectrics. While the classical principles of CR are well understood, possible quantum effects were only recently discussed. Interestingly, quantum mechanical aspects of CR address the following questions: (i)- Is it the matter or the electron which emits the photon [74]? (ii)- Which form of the photon momentum is to be exploited, the Minkowski or the Abraham representation [76]? Both these issues are at the heart of quantum electrodynamics. In most practical situations, the classical treatment which excludes the second quantization of the radiation agrees with the experimental results [73]. It is only at the extreme regimes of ultraslow electrons and ultrahigh photon energies, where quantum-optical corrections need to be considered [64]. However, when an electron wave packet which sustains certain spin and angular momenta is considered, novel aspects such as the splitting of the CR cone into two cones and radiation in the backward direction have been discussed [72]. Hereafter, only the classical treatments are considered.

CR emission can be observed in a transmission electron microscope as a result of the interaction of the electron beam with the sample. CR leads to a resonance in electron energy-loss spectra from various materials [77, 78], and the dispersion diagram associated to CR can be retrieved using momentum-resolved EELS [79]. The analytical expression for electron energy-loss spectra associated with CR in a bulk material has been provided by Kröger [66]. To enable a direct application to metamaterials, we have calculated the EEL spectrum using a vector potential approach as described in the appendix. The momentum-resolved EEL spectrum and the radiated power originating from the interaction of a moving electron with speed $V$ along the $z$-axis within a bulk material can be generalized for the case of an anisotropic and permeable material as:

$$P_z(\omega; k_x, k_y) = \frac{e^2}{8\pi^2 \varepsilon_0 V} \text{Re}\left(\frac{k_x^2}{\varepsilon_{rxx}} + \frac{k_y^2}{\varepsilon_{ryy}}\right) \times \frac{1}{\left|\varepsilon_{rzz}\mu_r k_0^2 - k_x^2 - k_y^2 - (\omega/V)^2\right|^2}$$

(3),

and

$$\frac{d\Gamma^{\text{EELS}}(\omega; k_x, k_y)}{dz} = \frac{e^2}{4\pi^2 \hbar \omega^2 \varepsilon_0} \times \text{Im}\left\{\frac{1}{\varepsilon_{rzz}} \frac{\left(\varepsilon_{rzz}\mu_r k_0^2 - (\omega/V)^2\right)}{\varepsilon_{rzz}\mu_r k_0^2 - k_x^2 - k_y^2 - (\omega/V)^2}\right\}$$

(4)

, respectively. For the sake of simplicity, the only nonzero components of the permittivity tensor are assumed to be $\varepsilon_{rxx}$, $\varepsilon_{ryy}$, and $\varepsilon_{rzz}$. Moreover, $\vec{k} = (k_x, k_y, k_z = \omega/c)$ is the wave vector of the emitted light in the reciprocal space, $e$ is the electron charge, $k_0 = \omega/c$ and $\varepsilon_0$ are the free-space wavenumber and permittivity, respectively. Interestingly, the EEL spectrum for a bulk material is only related to the $\varepsilon_{rzz}$ relative permittivity component, although for the emitted power in the z-direction all permittivity components are relevant. Moreover, the EEL spectrum has two contributions from the longitudinal and transverse terms given by

$$\frac{d\Gamma^{\text{EELS}}(\omega; k_x, k_y)}{dz} = \frac{e^2}{4\pi^2 \hbar \omega^2 \varepsilon_0} \operatorname{Im} \frac{1}{\varepsilon_{rzz}}$$
$$+ \frac{e^2}{4\pi^2 \hbar \omega^2 \varepsilon_0} \operatorname{Im} \frac{1}{\varepsilon_{rzz}} \frac{k_x^2 + k_y^2}{\varepsilon_{rzz} \mu_r k_0^2 - k_x^2 - k_y^2 - (\omega/V)^2} \quad (5)$$

where the first term in the right side of eq. (5) is known as the longitudinal term and the second term as the transverse term. This nomenclature is because of the longitudinal and transverse current densities in a bulk material both of which contribute to the energy loss experienced by the electron [80]. Examples for longitudinal excitations are the bulk plasmon, whereas for transverse excitations CR. As EEL spectra are routinely employed to perform Kramers-Kronig analysis, care should be taken regarding the transverse part [79], which is also theoretically confirmed by using eq. (5) (figure 2).

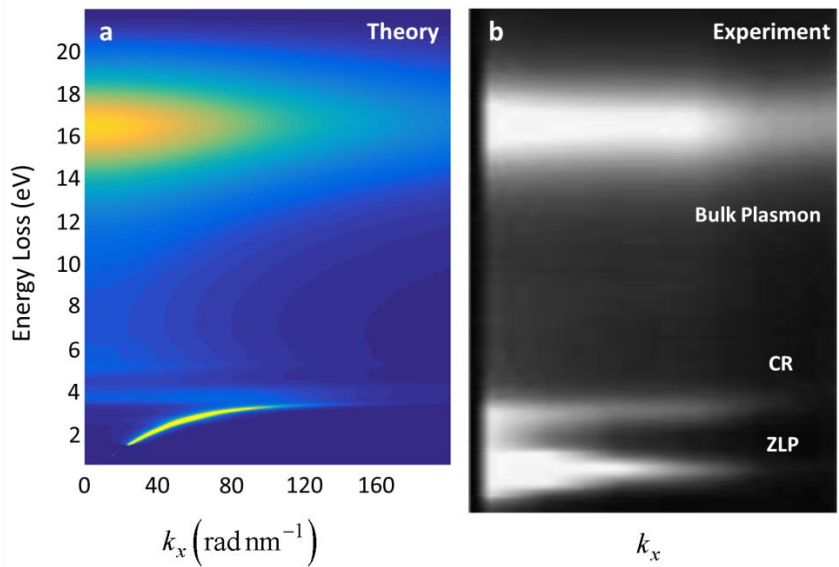

**Figure 2.** Momentum-resolved EEL spectrum for bulk silicon, for which the CR and bulk plasmon contributions are apparent. (a) theoretical EEL spectrum calculated using eq. (4). (b) Experimental EEL spectrum reported by Stöger-Pollach et al (without scale). Reproduced with permission from ref. [79], © Elsevier Ltd.

For anisotropic materials, direct detection of the CR emission by means of angle-resolved CL spectroscopy [20, 81, 82] can unravel the energy–momentum isofrequency surfaces. The isofrequency surfaces for anisotropic dielectrics are ellipsoids, whereas for hyperbolic metamaterials they are hyperbola. An example of a natural hyperbolic metamaterial in the visible frequency range is $Bi_2Se_3$ [83, 84], a uniaxial material with two different permittivity components (figures 3(a) and (b)). The structure of $Bi_2Se_3$ is composed of stacks of quintuple layers as shown in the inset of Figure 3a, where the permittivity within the *ab* crystallographic plane is different from the permittivity normal to this plane. We consider two different orientations of electron trajectories with respect to the optic axis in $Bi_2Se_3$; i.e., parallel (Figure 3(c)) and normal (Figure 3(d)) to the optic axis. The kinetic energy of the electron is assumed as 200 keV. Figure 4 shows the momentum-resolved EEL and also the power spectra for bulk $Bi_2Se_3$, using the permittivity data shown in Figure 3(a) and (b) [84]. For an electron traveling along the *c*-axis, as shown in Figure 3(c), there is a clear CR emission at energies below $E = 1.1$ eV at which the material is dielectric. This emission is apparent in both the EEL and in the radiated power spectra. Moreover, the emitted light has momentum and phase velocity components parallel to the electron velocity.

The excitation of the longitudinal (bulk) plasmon is also visible in the EELS data. Moreover, for energies above $E = 3.5$ eV, a reversed CR emission can occur in $Bi_2Se_3$ (figure 4(b)). A reversed CR emission occurs when emitted light propagates backward with respect to the direction of propagation of the electron. As the electron here propagates along the *z*-axis, a reversed CR leads to negative values for the power flow of the emitted light (Poynting vector) projected along the *z*-axis (eq. (3)). The momentum-resolved power spectrum shown in Figure 4c demonstrates the formation of spheroidal isofrequency surfaces at the plane normal to the electron trajectory, because for this case we have $\varepsilon_{r\,xx} = \varepsilon_{r\,yy} = \varepsilon_{r\,a,b}$.

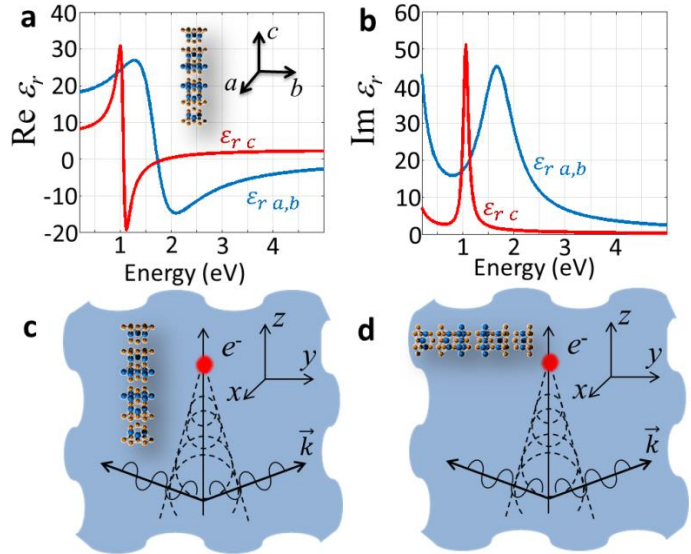

**Figure 3.** Relative permittivity components of $Bi_2Se_3$. (a) real part of the permittivity, and (b) imaginary part of the permittivity components along *c*-axis (red lines) and normal to it (*a-b* plane). Two cases of electron trajectories are considered, (c) electron traveling normal to the optic axis, and (d) electron traveling parallel to the optic axis.

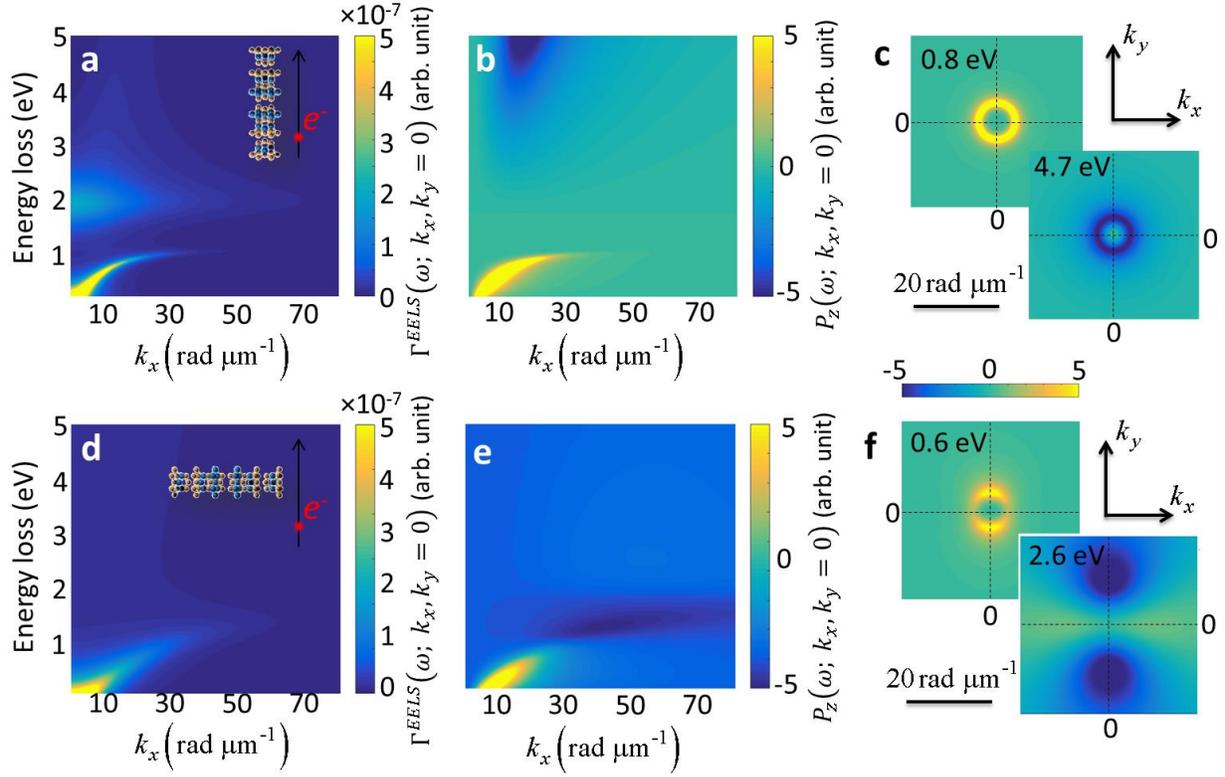

**Figure 4.** (a) EEL and (b) radiated power spectra for an electron at a kinetic energy of 200 keV traveling parallel to the *c*-axis (the configuration is displayed in figure 3c). (c) The momentum distribution of the emitted light at two different photon energies depicted at each frame. (d) EEL and (e) radiated power spectra for an electron at the kinetic energy of 200 keV traveling parallel to the *b*-axis (for configuration see figure 3d). (f) The momentum distribution of the emitted light at two different photon energies depicted at each frame.

The situation is quite different when the electron is traveling parallel to the optic axis (the *c*-axis, see figure 3 (a)). For this case, although the CR emission is still apparent at lower energies, there is no bulk plasmon contribution within the energy range investigated here despite the fact that $\varepsilon_{r\,a,b}(\hbar\omega = 1.72\,\text{eV}) = 0$. This is due to the fact that the loss function $\text{Im}(1/\varepsilon_{r\,a,b})$ does not exhibit any maximum at this energy. The radiated power spectrum exhibits a reversed CR emission at energies between 1.06 eV and 1.80 eV, for which $\varepsilon_{rc} \leq 0$ but $\varepsilon_{r\,a,b} > 0$ holds. The momentum-resolved power spectrum clearly certifies the anisotropic nature of the material. For the conditions $\varepsilon_{r\,a,b} > 0$ and also $\varepsilon_{rc} > 0$, an ellipsoidal ring should appear in the momentum-resolved power spectrum as shown in Figure 4f for the panel resolved at the energy of $E = 0.6$ eV. At energies higher than $E = 1.8$ eV at which $Bi_2Se_3$ is hyperbolic, hyperbolic-like features appear, instead of the ellipsoidal ring.

A left-handed material (LHM) is a material whose permittivity and permeability and hence refractive index have simultaneously negative values, at least in a certain frequency range [85]. Reversed CR emission has been first assured for a charge particle moving inside a LHM [86]. The observation of anomalous maxima at field intensities inside the LHM for an electron crossing an interface between LHM and air is related to the phase velocity of the CR emission in the LHM which is antiparallel to the group velocity [87]. The momentum-resolved EEL and CL spectra calculated for an electron propagating inside a bulk LHM directly demonstrate the negative phase velocity of light in the LHM, and correspondingly the reversed CR emission [88]. In order to confirm this, we consider here a prototype isotropic LHM with permittivity and permeability components both given by a Drude model as $\varepsilon_r = \mu_r = 1 - \omega_p^2 / \omega(\omega - i\gamma_p)$, where $\omega_p$ is the plasma frequency and $\gamma_p$ is the damping rate. Using eqs. (2) and (3), the momentum resolved EEL and radiated power spectrum are calculated, and the results are shown in Figure 5. The calculated power flow (pointing vector) demonstrates negative values, which is due to the antiparallel propagation of the emitted light with respect to the direction of propagation of the electron. In other words, the negative refractive index leads to negative values for power and hence a backward emission up to the plasma frequency (figure 5(b)).

Finally, it was recently demonstrated that the interaction of a moving charge with a hyperbolic material can lead to the CR emission without any required threshold for the electron velocity [89]. This finding has been further used to employ low-energy electrons and hyperbolic materials in an integrated setup to generate a tunable and compact light source.

*2.4. Transition radiation*

A moving electron crossing an interface between two different materials generates a radiation called transition radiation (TR), which is mutually coherent with the self-field of the electron. It is widely accepted that TR happens because of the interaction of the moving electron with its image charge. This involves (i) the creation of an image charge in the second medium when the electron is still moving in the first medium, (ii) a gradual annihilation of the image when the electron approaches the interface, and (iii) the screening of the electron charge when the electron has crossed the interface. This effect has been first predicted by Ginzburg and Frank [90], and thereafter intensively studied both theoretically and experimentally by many groups. Ginzburg considered several ways to treat the problem theoretically either as a Larmor radiation created due to a *sharp* change in the velocity of the electron at the interface [91], or a sudden change of the refractive index and hence of the phase velocity of the light waves due to the inhomogeneity of the medium [92]. Interestingly, TR can be also considered within the category of radiation by a uniformly moving charge, and therefore a theoretical treatment based on the non-recoil approximation is well suited to describe this phenomenon. Just like CR, TR also contributes to the EEL spectra. A detailed analytical treatment of EEL probability has been performed by Kröger already in 1968 [66], and more recently by Garcia de Abajo et al. [93] in more detail considering inclined excitations towards the surface. Taking into account the retardation effect and boundary conditions, such a treatment has been shown to be in a good agreement with experimental results [93].

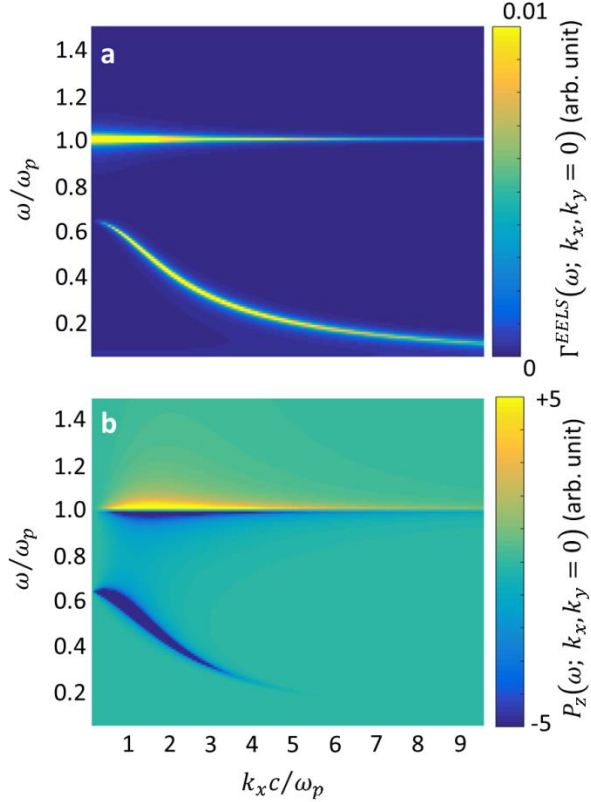

**Figure 5.** Calculated (a) EEL and (b) power spectra for an electron at the kinetic energy of 200 keV propagating inside a LHM, with both the permittivity and permeability values modelled by a Drude function.

Coherent TR in the far-infrared could be detected and carefully configured from the overall CR and TR radiation pattern, by carefully tuning the refractive index of the materials involved [94]. Using TR emission, Glinec et al. were able to resolve spatiotemporal fine structures of laser-driven electron beams in the sub micrometer range [95]. In relevant works, the longitudinal micro-bunching of the electron beams in a free electron laser has been detected using coherent TR from the interaction of electron beams with thick foils [96-98]. Moreover, TR has found application in developing intense THz radiation sources [99, 100].

Although TR takes place already at a single interface, practical situations consider more often thin films. In thin films, TR, CR, and also guided modes contribute significantly to the radiation loss and energy loss suffered by the electron. Importantly, in isotropic metallic thin films below the plasma frequency, CR cannot take place. However, in this energy range surface plasmons contribute significantly to the EEL and also to radiation spectra. It is known that far-field optical radiation cannot excite bulk plasmons, and accordingly longitudinal currents and hence bulk plasmons in a reciprocal picture cannot contribute to the far-field radiation as well. However, the situation changes when an emitter like a relativistic electron is placed inside matter [26]. In this case, EELS have already demonstrated a significant signal due to the excitation of bulk plasmons by moving electrons [101], which can be exploited to map different elements of composites.

Inclusion of a second interface and also of multilayered structures has another consequence, namely the occurrence of interferences within the thin film, because of multiple reflections of the emitted light from the boundaries (figure 6(a)). This interference phenomenon affects the overall EELS signal and also the TR emission significantly, and can be used to measure the electron beam energy with high accuracy [102]. Moreover, dielectric and metallic thin films are also able to support guided modes, and the field profile in the latter case can be decomposed into symmetric and antisymmetric modal configurations formed by the coupled surface plasmon polaritons at the interfaces. The hybridization of surface plasmon modes for pure thin films is perfectly captured in the momentum-resolved EEL spectra [103]. For hyperbolic metamaterials like $Bi_2Se_3$, there exists a more complicated classification of optical modes [83, 104]. In general, six different modal classes are possible, specifically symmetric and antisymmetric transverse magnetic with respect to $x$, $y$, or $z$ axes ($TM_x$, $TM_y$, and $TM_z$, respectively), where part of which can be excited by moving electrons interacting with thin films composed of hyperbolic materials. Moreover, as a result of the interaction of electron with a $Bi_2Se_3$ thin film, an enhanced ultrabroadband TR will be generated (figure 6(b)), which manifests itself in ultrabroadband EEL resonances as well [83]. This is due to the hyperbolic nature of $Bi_2Se_3$ which provides a very high photonic local density of states [105]. In fact, hyperbolic metamaterials have been already demonstrated to be able to enhance and control the emission from nanoemitters in an efficient and directional way [106]. However, the possibility of controlling the directionality of TR with hyperbolic materials has yet to be fully explored.

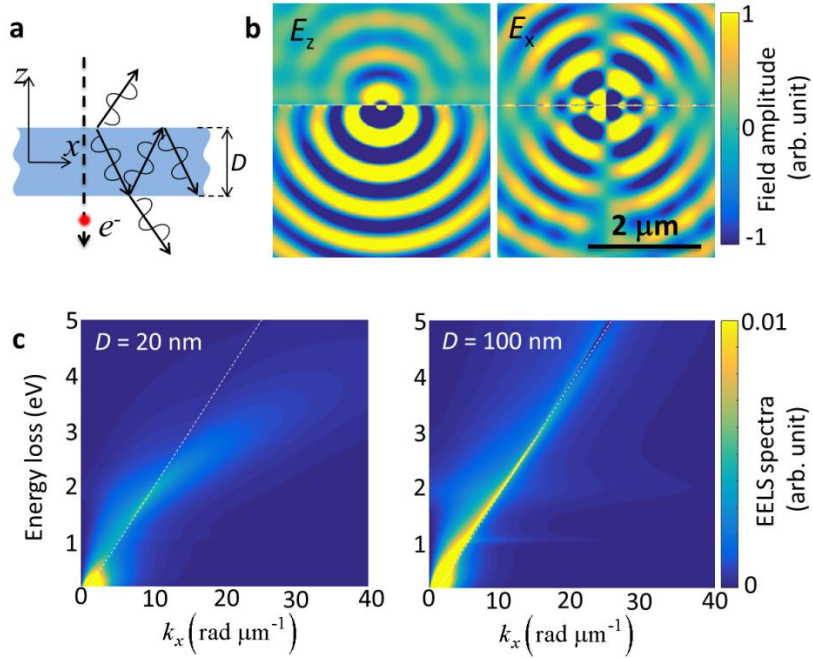

**Figure 6.** (a) A moving electron interacting with a thin film can excite TR and CR, as well as guided modes formed by multiple reflections within film. (b) Field profile of the excited states at the energy of 2 eV from an electron at a kinetic energy of 200 keV interacting with a $Bi_2Se_3$ film with a thickness of 20 nm. The optic axis is parallel to the film. (c) Momentum-resolved EEL spectra for a 200-keV electron interacting with $Bi_2Se_3$ films with thicknesses depicted in each frame.

As previously mentioned, EELS can map the photonic local density of states projected along the electron trajectory [4]. This particular projection along with the momentum conservation picture sets a criterion for the EELS to be able to map only certain modes of the structures [21]. Nevertheless, the mentioned criterion is advantageous in designing coherent radiation sources using synchronization, as in Smith–Purcell free-electron lasers [107, 108] (see section 2.4). Here, we only mention that using EELS only those modes of the structure are excitable which support an electric field component along the electron trajectory. In other words, $TM_y$ modes are not excited as these modes sustain only $E_y$, $H_x$, and $H_z$ field components. Nevertheless, the dispersion of the excitable modes, can be captured using momentum-resolved EELS (figure 6(c)). For a thin film with a thickness below 20 nm, the dispersion of the $TM_x$ mode is clearly visible, whereas for a film as thick as 100 nm, the contributions of CR and bulk plasmon are dominant.

Interaction of extremely concise wave packets of relativistic electrons with thin samples and interfaces takes place within only a few attoseconds. This short interaction time provides a large bandwidth for the spectroscopic investigations of optical modes, and naturally leads to ultrashort light emission, which has been recently studied by taking the Fourier transform of time-harmonic solutions of the Maxwell equation over a large spectral bandwidth [109].

Finally, it should be emphasized that there still exists a great potential for controlling the emission from electron-driven photon sources using metamaterials. The above-mentioned example was including a natural hyperbolic metamaterial. An even more interesting phenomenon is to be exploited by using LHM. As a prototype, a LHM film with both permittivity and permeability modeled by a Drude function as $\varepsilon_r = \mu_r = 1 - \omega_p^2 / \omega(\omega - i\gamma_p)$ is considered. Figure 7(a) displays the calculated momentum-resolved EEL spectra, considering an electron at a kinetic energy of 200 keV interacting with a film with the thickness $D = 2c/\omega_p$. The exotic dispersion of the possible excitations is well captured in this picture. Inverted CR emission with the specifications described in section 2.2, along with the excitation of SPP modes which are decomposed into the symmetric and antisymmetric modes take place, which demonstrate a weak interaction with each other. Bulk plasmon excitation exactly at the plasma energy, as well as a plethora of guided mode excitations occurring at energies below the CR dispersion are also inferred from the momentum-resolved EEL spectrum. The different crossings which happen for CR and SPPs can be an interesting option for investigating weak and strong couplings in metamaterials [110, 111], TR emission from such a structure reveals the interferences between the inverted CR and TR, which can be understood based upon the splitting of the field profile at $z > D/2$, in contrast to the transmitted field (figure 7 (b)).

*2.5. Smith–Purcell Effect*

Already in 1953 [112], even before detection of TR by Goldsmith and Jelly in 1959 [113], Smith and Purcell investigated the emission properties of an electron beam propagating parallel to a metal diffraction grating. Based on geometrical-optics assumptions, they found that there is a simple relation between the wavelength of the emitted light and the velocity of the electron as $\lambda = d\left(\beta^{-1} - \cos(\theta)\right)$, where $d$ is the period of the grating, $\beta = v/c$, and $\theta$ is the angle between the emitted light ray and the velocity of the electron ($v$). In fact, this relation can be rationalized by the momentum conservation criterion given by eq. (2), by considering $k = k_0 + 2m\pi/d$, where $m$ is the diffraction order. First harmonic Smith–Purcell radiation occurs whenever $m = 1$ is the dominant response.

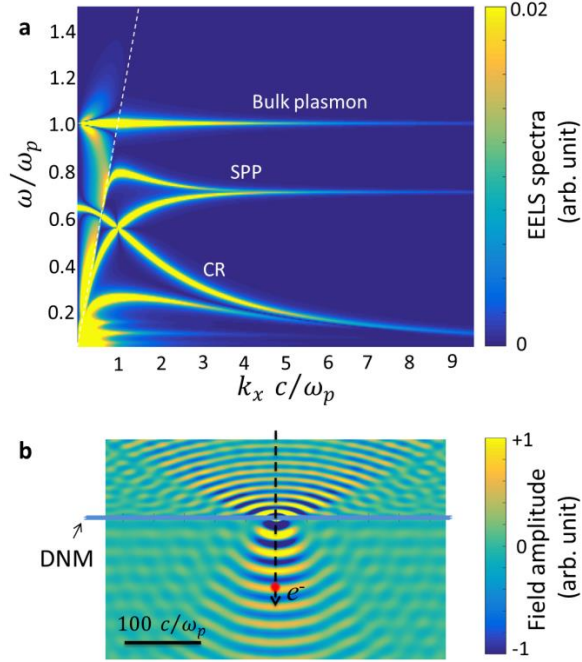

**Figure 7.** (a) Momentum-resolved EEL spectra for an electron at the kinetic energy of 200 keV interacting with a LHM film with the thickness of $D = 2c/\omega_p$. (b) Profile of the $z$-component of the electric field at $\omega = 0.62\omega_p$.

Although Smith–Purcell emission is generally described by the grating diffraction orders, practical situations typically involve a finite number of grating elements. The effect of the number of grating elements on the probability of photon generation and EELS has been addressed in ref. [114], which demonstrates a transition from individual resonances to an interference pattern in the far-field upon increasing the number of elements (figure 8(a)). Moreover, it has been shown that for the case of a grating composed of silica spheres, the EELS and photon generation probabilities are quite similar, in contrast to metallic gratings. The interplay between the Smith–Purcell radiation and surface plasmon radiation caused by the interaction of an electron beam and a metallic grating has been recently investigated experimentally, using a silver grating, where it has been found that the interferences between plasmons and Smith–Purcell radiation lead to a characteristic Fano resonance (figure 8(b)) [20]. In a relevant contribution, using the Poynting theorem, it has become possible to derive a relation between the photon generation probability and EEL probability [115]: The difference between EEL and photon generation probabilities is related to the absorption spectra, which in turn is governed by the dissipation loss in the matter.

Considering all the mechanisms of radiation described above, whether the emission is coherent or not depends not only on the material, but also on the shape of the electron wave packet [116]. As an example, for a bunched electron wave packet, the emission is coherent when the longitudinal broadening of the bunch is smaller than the wavelength of the emission [117]. In general, an ensemble of emitters positioned in such a bunch can lead to a coherent spontaneous emission known as superradiance [118]. Interestingly, this phenomenon is time-reversable, as a coherent optical radiation interacting with an electron wave function can cause the wave function to evolve into bunches [119, 120], even for a single electron wave packet [121].

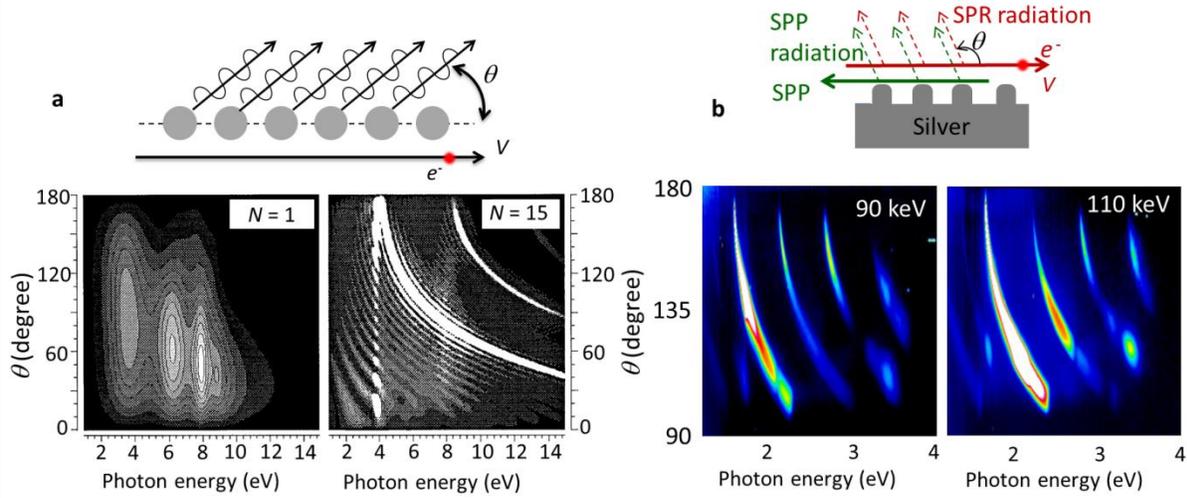

**Figure 8.** (a) Simulated probability of photon generation from the interaction of a relativistic electron at the kinetic energy of 200 keV with a chain of Al spheres composed of $N = 1$ and $N = 15$ elements. Reproduced with permission from [114], © American Physical Society. (b) Measured CL spectra from the interaction of an electron beam at different energies with a silver grating. Reproduced with permission from [20], © American Physical Society.

### 3. Electron-driven coherent radiation sources for the visible and UV range

#### 3.1. Smith–Purcell photon sources

The possibility to tune the wavelength of the emitted light by either changing the period of the grating or the velocity of the electron, has made the Smith–Purcell effect one of the prominent choices for designing THz radiation sources [122, 123], in addition to superconductors [124]. In general, the Smith–Purcell effect is routinely used to design tunable electromagnetic sources in frequency and power ranges which are not directly accessible by semiconductor-based lasers [107, 108, 125-135]. Smith–Purcell radiation is one of the different mechanisms for designing free electron lasers, besides bremsstrahlung and Thomson scattering [136]. Here the effort is to maximize the emission gain per electron bunch and to enhance coherence, by introducing well-controlled electron beams as well as feedback elements [137-139]. Whereas the bremsstrahlung radiation is mostly used for generation of X-rays at ultrahigh intensities,

Smith–Purcell emission is used for generating table-top and compact THz radiation sources, for instance in the form of devices called Orotrons [140].

Major advances in the development of Orotrons include (i) Improvement of starting-current-density threshold and radiation power by incorporating multimode slow-wave structures, composed of the grating and a flat mirror, which is designed to optimize the coherent waves in the form of s-polarization [141]; (ii) Engineering the shape of the electron wave function as a pre-step, in order to generate bunched electron beams, which upon interaction with a grating can produce a coherent radiation at the same frequency of the imposed laser excitation [142]; (iii) Considering aperiodic chains of nanospheres in the form of tessellated and Fibonacci patterns; (iv) Including Babinet metasurfaces which offer a better control over the polarization state of the emitted light in comparison with metallic gratings [143]; (v) Inclusion of a compact cylindrical optical corrugated waveguide and a feedback mirror, which offers an enhanced Smith–Purcell radiation and high photon generation efficiency [144]; (vi)

## 3.2. Metamaterial-based electron-beam-driven photon sources

In distinction to the Smith-Purcell emission, here photon generation mechanisms based on TR, CR, and diffraction radiation is considered. In an interesting configuration (figure 9(a)), Bendana et al. demonstrated that an electron beam interacting with an optical fiber can generate a single photon source which is then coupled to the fundamental mode of the fiber in a near-deterministic way , which is captured in the calculated photon yield (figure 9(b)) [145]. It has been subsequently shown that the composition of nanoantennas coupled to a plasmonic rib waveguide, can be used for controlling the recoil that the electron receives and also to generate single photons in a unidirectional emission pattern [115].

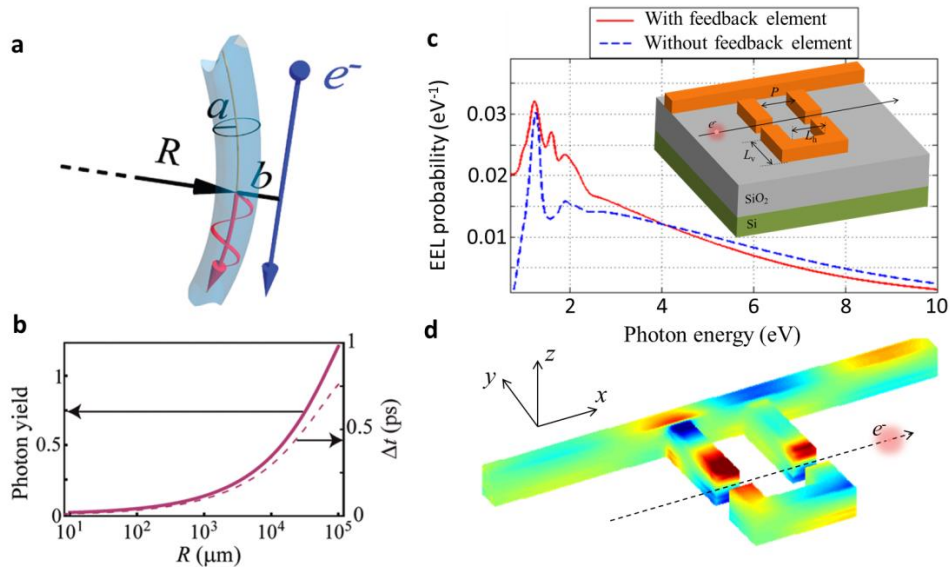

**Figure 9.** (a) A single electron interacting with an optical fiber can generate single photons with high efficiency for large values of the fiber radius. (b) Yield of the generated photon which is coupled to the fundamental mode of the waveguide. Reproduced with permission from [145]. © American Chemical Society. A planar configuration of nanoantennas coupled to a resonator and a plasmonic rib wave guide can be used to generate photons which then propagate in the waveguide in a unidirection way. (c) EEL probability with and without including the resonator (feedback element), and (d) the calculated $z$-component of the electric field. Adapted from [115].

Zheludev and coworkers showed that a tunable light source can be configured by the interaction of electron sources with gold/silica multilayer structures (figure 10(a)) [146, 147]. Whereas there are some similarities between this configuration and that of a Smith–Purcell free electron laser, the configuration used by Adamo can be used to generate broadband emission in the visible range. Nevertheless, the emission from their structure was expected to be incoherent. In another contribution, Zheludev and coworkers demonstrated that a metamaterial-based planar structure traversed by an electron beam emits photons which are coherent with the near-field of the electron (figure 10(b)), which is mainly due to the collective excitation of the metamaterial elements and the excitation of surface plasmons [148]. Moreover, when the electron beam is focused on the metamaterial elements instead of the solid film, the emission is further enhanced. Recently, it has been demonstrated by the same group that holographic patterns can be exploited to control the directionality, the wavelength, and the polarization state of the generated light in a fully coherent way [149] (Figure 10(c)). Interestingly, generation of optical vortex beams with topological charges of up to 10 has been demonstrated by such holographic patterns. It has been concluded that holographic structures in comparison with thin films can enhance the brightness by up to two orders of magnitude.

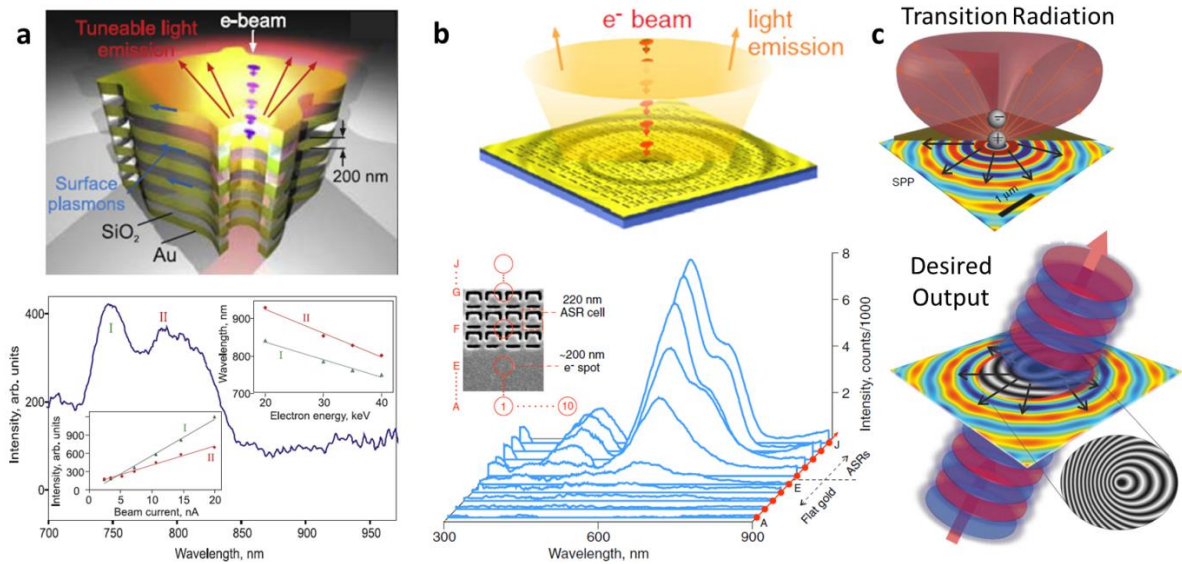

**Figure 10.** (a) Upper frame: Topology of a tunable 'light-well' electron driven source. Lower frame: Spectrum of the reflected light from such a configuration. The intensity of the emitted light versus electron beam current is shown in the lower inset, and the wavelength of the emitted light versus the electron energy is depicted in the upper inset. Reproduced with permission from [147], © IoP publishing. (b) Upper frame: Configuration of the electron-beam-driven metamaterial light source. Lower frame: the spectrum of the emitted light for different electron impacts shown in the inset. Reproduced with permission from [148], © American Physical Society. (c) Upper frame: normal transition radiation for an electron traversing a thin film. Lower frame: Transition radiation from a holographic mask. Adapted from [149].

Another strategy to control the directionality of the emission and at the same time to enhance the radiation is a concomitant utilization of hyperbolic metamaterials and photonic crystals [150]. Whereas the hyperbolic materials serves to provide a better coupling efficiency of the electron-induced polarization to the radiation continuum [83], the photonic crystal structure helps to control the directionality of the emission and also to enhance the Purcell effect, which also enhances radiation. As an example, an inverted super lens composed of $Ag/Al_2O_3$ multilayers and an incorporated hexagonal photonic crystal can be used to generate a unidirectional radiation pattern along the electron trajectory, in the form of ultrashort optical transverse-magnetic Laguerre–Gaussian pulses.

*3.3. Applications of electron-driven photon sources*

*3.3.1 Quantum-optical experiments*

Future quantum computing and information technologies necessitate gaining control over the generation, propagation, and detection of single photons. Usually single photon states in vacancies and semiconductor-based nano-circuitries are controlled and detected by means of optical methods like photoluminescence. Tizei and Kociak demonstrated for the first time that electron probes can be used to generate such photon states in nitrogen-vacancy centers on the nanometer scale (figure 11(a)) [151]. By combining the CL detection system with a Hanbury Brown–Twiss intensity interferometer, the second-order correlation function could be measured, which showed a dip at few nanosecond temporal delays, which is a clear signature of photon antibunching (figure 11(b)). Moreover, electron-induced single-photon states in individual CdSe/CdS quantum dots were studied and directly compared to the photoluminescence spectra (figures 11(c) and (d)) [152]. Furthermore, CL and photoluminescence of ensembles of defect centers of diamond and hexagonal boron nitride were studied and compared by measuring the second-order auto-correlation function [153, 154].

*3.3.2. Time-resolved spectroscopy*

Besides the direct excitation of single photon sources like quantum dots and defect centers, electron-driven metamaterial-based photon sources are suitable for performing spectral interferometry within the electron microscopes, as conjectured recently by means of simulations [150]. This is achieved by exposing focused transition radiation of the electron, which is mutually coherent with the near-field of the electron self-field, onto a sample (figure 12(a)). The sample is located at the distance $L$ from the electron-driven photon source (EDPHS) (figure 12(b)). Because the electron and photons propagate at different velocities in the vacuum, the delay between the electron beam and the TR at the sample position is controlled by changing the distance $L$ as $\delta\tau = (1-\beta)L/V$, where $L$ is the speed of the electron, $\beta = V/c$, and $c$ is the vacuum speed of light. Moreover, by changing the distance, the relative phase of the electron-induced and photon-induced excitations in the sample can be further controlled, which causes interference patterns in the acquired EEL [155] or CL spectra (figure 12(c)). The acquired EEL or CL energy–distance map allows one to extract the spectral phase of the electron-induced excitations relative to the photon-induced excitations [150].

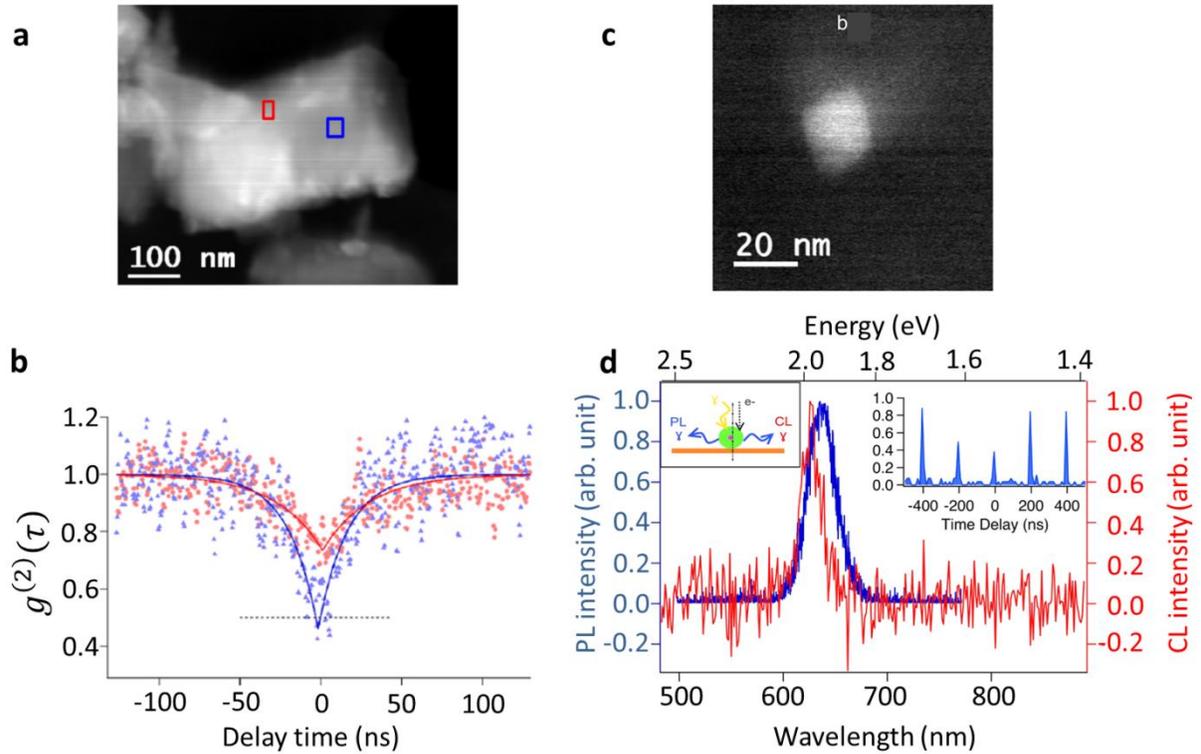

**Figure 11.** (a) Annular dark-field image of a diamond nanoparticle. (b) Second-order auto-correlation function acquired for different positions of the electron beam, depicted at the image. (a) and (b) panels are reproduced with permission from [151]. © American Physical Society. (c) high-angle annular dark-field image of a single CdSe/CdS quantum dot. (d) CL and photoluminescence spectra of the same quantum dot. Left inset: schematic of the experiment. Right inset: photoluminescence correlation time. Reprinted with permission from [152], © American Physical Chemistry.

## 4. Mesoscopic metallic and dielectric laser accelerators

Linear acceleration, either with dielectric or metallic gratings, is a separate field of study. Extensive publications and reviews have already been devoted to this topic, with an ongoing effort to optimize and miniaturize the accelerators [156]. Here, the focus is on the peculiar similarities between the two branches of science, namely electron-based radiation and linear acceleration, and possible applications in electron-beam shaping. As mentioned above, synchronization between a grating diffraction mode and the electron beam leads to a coherent Smith–Purcell radiation. In an inverse approach, illuminating the grating with an intense laser radiation will cause the electrons travelling adjacent to the grating to gain energy. This effect has been first proposed by Shimoda in 1962 [157] and experimentally realized by Mizuno et al. 25 years later [158]. The grating acts as a mediator for transferring the energy and momentum of the light to the electron beam. Depending on the incorporated grating, the mesoscopic accelerators may be divided into three categories, namely plasmon-, metamaterial-, and dielectric-based linear accelerators (LINACs).

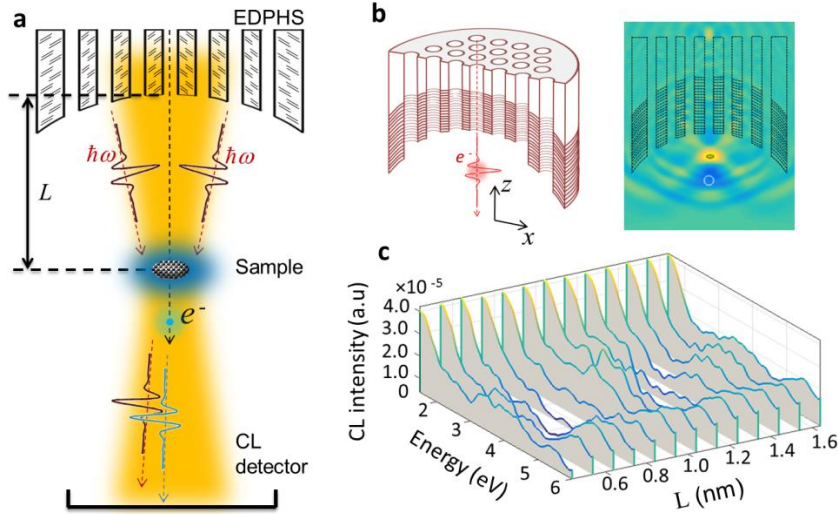

**Figure 12.** (a) Schematic of the proposed spectral interferometry experiment with an electron microscope. (b) an example of an electron-driven photon source. Right panel shows the *z*-component of the electric field radiated from the EDPHS, at a certain time. (c) Calculated CL energy–distance map versus energy and distance *L* between the sample and the EDPHS. Adapted from [150].

*4.1. Plasmon- and metamaterial-based linear accelerators*

Although the acceleration based on the interaction of electrons with a laser-illuminated plasma has been used to accelerate particles [159], plasmon-based accelerators can offer a more concise mesoscopic acceleration device. The first experiments regarding the inverse Smith–Purcell effect incorporated metallic gratings [158, 160]. This concept has been confirmed again in 2003 by studying both low-energy and relativistic electrons [161]. Recently, it has been demonstrated that the near-field of an electron beam can also be amplified by synchronizing the electron with the evanescent tail of the laser-induced surface plasmon mode [162], in well-designed configuration positioned inside a scanning electron microscope. Even more concise acceleration platforms may be practically realized by considering the concepts in designing metamaterials for improving the stability of the beam as an example [163].

*4.2. Dielectric-based linear accelerators*

Despite the enormous near-field enhancement of up to several orders of magnitude which occurs near metallic gratings because of the excitation of surface plasmons [164], the measured acceleration gradient for metallic-based LINAC is about two orders of magnitude smaller than for the RF linac. This difference arises because of the low radiation damage threshold of metals. In this regard, dielectric-based accelerators with higher damage threshold are viable for higher acceleration gradients [165]. Dielectric LINACs paired with laser fibers can provide a miniaturized accelerating platform [166, 167]. Additionally, low energy photoemission electrons from needle tips can be further accelerated using dielectric LINACs, where the combination of two structures can form a miniaturized electron gun which generates coherent pulsed electrons at desired energies [168].

In general an efficient and stable beam accelerator must fulfill the following requirements: (i) The synchronicity condition has to be considered in order to provide a continuous transfer of the momentum from the laser illumination to the electron beam by means of the near-field distribution of the grating [169-171]. (ii) Feedback elements are routinely included in the design of Smith–Purcell free-electron lasers, and can also improve the efficiency of the dielectric LINACs. (iii) During the course of interaction with the laser beam and gratings, the particles are accelerated in a quite fast way, and soon become out of phase with the synchronize mode. This phenomenon is called dephasing. In order to account for this change, multiple stages of accelerators and chirped gratings may be used [172-174]. (iv) Only longitudinal forces originating from the longitudinal components of the electric field are helpful for acceleration. However, the near-field distribution of the grating enforces also a transverse force, which is out of phase with the longitudinal force, and causes defocusing of the electron beam. In order to avoid defocusing, a symmetric configuration needs to be implemented [175]. Indeed an ideal platform might be the closed form waveguide positioned inside a 3D photonic crystal [176]. However, due to the formation of a quasi-cavity in the symmetric configurations, near-field-enhanced Kapitza–Dirac-like diffraction of electron wave functions is another source of defocusing [121].

## 5. Shaping the electron wave function

The nonrelativistic solutions for the wave equation in a cylindrical system are given by a set of complete basis functions defined as $\psi(\rho,\varphi,z;\omega) \alpha \; \exp(ik_z z) \, B_m(k_\rho \rho) \exp(im\varphi) \times \exp(-i\omega t)$, where $B_m$ is the Bessel function of order $m$, $k_z^2 + k_\rho^2 = 2m\omega/\hbar$, $\hbar\omega$ is the kinetic energy of the electron, and $m$ is the azimuthal order. On this basis, evolutions of waves in cylindrical systems are well described by three parameters which define degrees of freedom of beams, namely $m$, $k_\rho$ or $k_z$, and the spin of the electron defined by $s = \pm 1/2$. In fact, it has been known for long time that an electron has a magnetic moment [177-179] associated with its spin, in contrast to photons. The interaction of the spin with an external magnetic field causes spectral fine structures of atoms and molecules [180]. Moreover, the electron has an orbital angular momentum. Specifically, the momentum density of an electron is given by

$$P = (\hbar/2i)\left(\psi^\dagger \vec{\nabla}\psi - (\vec{\nabla}\psi^\dagger)\psi\right) + (\hbar/4)\vec{\nabla} \times \left(\psi^\dagger \hat{\sigma}\psi\right) \quad (6)$$

in which $\hat{\sigma}$ is the Pauli matrix and $\psi$ is the electron wave function. The first and second terms in eq. (6) are associated with the linear momentum and spin, respectively. While the spin of an electron originates from the circulating flow of energy inside an electron wave function, the linear momentum is attributed to the electron in motion [181]. Moreover, linear momentum produces a coordinate-dependent orbital angular momentum, defined as $L = (\hbar/2i)\int \vec{r} \times \left(\psi^\dagger \vec{\nabla}\psi - (\vec{\nabla}\psi^\dagger)\psi\right)d^3\vec{r}$ [182]. Even free electrons can preserve an orbital angular momentum by having structured wave functions like Laguerre–Gaussian forms, which carry azimuthal phase structures. There is an intrinsic similarity between electrons and photons in this regard, as the statistical distribution of both in the time-harmonic representation is described by the Helmholtz wave equation [182]. The applicability of such a concept in controlling the magnetic moment of a free electron beam, which cannot be achieved for the spin-induced magnetic moment, and the fact that the effect of spin on elastic scattering is weak [183], together have opened a recent paradigm in electron microscopy.

Transversal electron beam shaping, however, covers a much broader concept beyond angular momentum manipulations [184, 185]. Realization of non-diffractive wave fronts [186] could offer a suitable means to tailor the electron trajectories in microscopes. In fact, in the ideal case non-diffractive wave patterns, such as Bessel beams, maintain their shapes and are immune to expansion via propagation in space. However, despite the great interest in optics towards diffraction-less photon beams, only little attention has recently been paid to electron Bessel beams [187, 188] and Airy beams [189].

The applicability of controlling the phase and shape of the electron wave function to decompose the plasmonic modes according to their symmetry has been demonstrated [190]. Another advanced way of shaping a quantum wave function has been used for several years to control quantum-coherent paths in solid-state physics [191] and chemistry [192], in order to manipulate the final atomic and molecular states towards a desired wave packet. In principle, it is also possible to employ free-space optical beams to coherently manipulate the shape of a swift electron wave function, as has been demonstrated already several years ago by the Kapitza–Dirac effect and also exploited experimentally [55]. Recently, exploiting the Kapitza–Dirac effect to generate electron beams with angular momentum has also been theoretically explored [193].

*5.1. Shaping the transverse components of the wave function with thin masks*

Electron vortices were first theoretically proposed by Bliokh et al. [194] and shortly after experimentally realized by Uchida and Tonomura by incorporating stacked graphite thin layers [195]. A relativistic electron passing through such a mask experiences different *optical paths* in regions with different thicknesses. At some specific positions of the sample, orientations of graphite stacks were such which created a singularity in the electron phase front, hence imposing an azimuthal order $m = 1$ over the electron. Soon afterwards, such masks were replaced by holograms [196, 197], which delivered more control over the azimuthal order $m$, to go beyond the value $m = 1$ (figure 13(a)). Currently, azimuthal orders up to $m=50$ [198] (figure 13(b) and (d)) and even $m = 400$ [199] (figure 13(c) and (e)) can be obtained by using nanofabricated computer-generated holograms. Electron vortex beams promise various applications in chemistry and solid-state physics [185], ranging from manipulation of nanoparticles by transferring the angular momentum to them [200] and energy-loss magnetic dichroism [201] to very fundamental aspects such as magnetic monopole moments sculptured in electron waves [202].

*5.2. Shaping the longitudinal components of the wave function with nanostructures*

While the above mentioned beam shaping involves a control of the transversal components of the electron beam into a superposition of angular momentum orders, more advanced manipulation of the shape of an electron wave function has to be achieved for an optimum control in three dimensions. An example for manipulating the shape of an electron wave function in the longitudinal direction is electron beam bunching [203] in which relativistic electrons interacting with magnetic undulators and plasma-based accelerators form bunches [204]. Interaction of pre-bunched electron beams with matter and optical gratings results in novel coherent radiation mechanisms like superradiant emission [205, 206]. The effects of pre-bunching and the modulation index of the electron beams on the wavelength and also the gain threshold of many different kinds of mechanisms of radiation in free-electron lasers like Cherenkov radiation [207, 208], surface-wave-enhanced radiation [208], travelling wave radiations [209], and masers [210] have received particular attention.

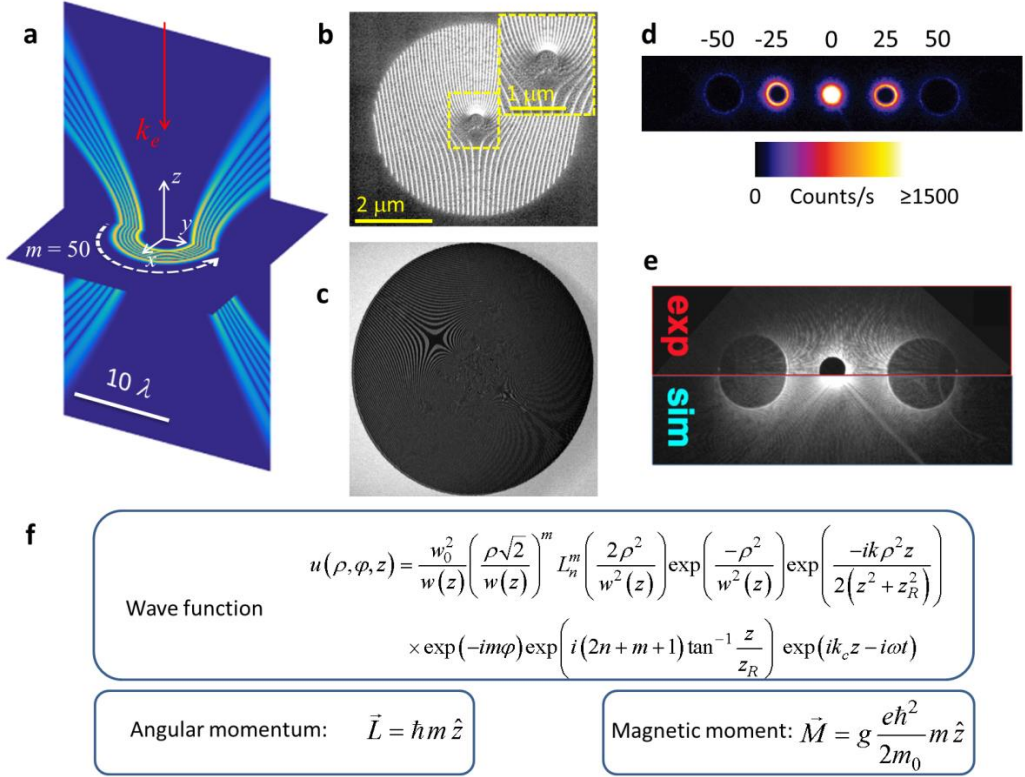

**Figure 13.** Electron vortex beams with controlled angular momentum orders. (a) Spatial profile of a Laguerre–Gaussian electron wave function for an electron with $m = 50$ and $n = 5$, at a kinetic energy of 50 keV. A hologram-generated mask for producing angular-momentum orders up to (b) $m = 50$, and (c) $m = 1000$. Electron intensity patterns detected after the interaction of the electron with the holograms shown in (b). (e) Experimentally observed and simulated patterns after the interaction of electron beams with the hologram shown in (c). (f) Spatial profile of the Laguerre–Gaussian electron wave function, angular momentum, and magnetic moment of a vortex electron beam in a cylindrical coordinate system designated by $(\rho, \varphi, z)$. Here, $n$, $m$ are integers, $w_0$ is the beam waist, $w(z) = w_0\sqrt{1+(z/z_R)^2}$, $z_R = \pi w_0^2/\lambda$ is the Rayleigh range, and $L_n^m(\cdot)$ are the Laguerre polynomials. (b) and (d) are reprinted with permission from [198]. © American Association for the Advancement of Science. (c) and (e) are reproduced with permission from [199]. © American Physical Society.

## 6. Self-consistent methods to simulate the interaction of electrons with light and matter

Quantum mechanical aspects of the interaction of electron with electromagnetic field can be described with the Schrödinger equation as

$$\left( -\frac{\hbar^2}{2m_0}\nabla^2 - \frac{i\hbar e}{m_0}\vec{A}(\vec{r},t)\cdot\vec{\nabla} + \frac{e^2}{2m_0}\left|\vec{A}(\vec{r},t)\right|^2 \right.$$
$$\left. -e\varphi(\vec{r},t) \right)\psi(\vec{r},t) = i\hbar\frac{\partial \psi(\vec{r},t)}{\partial t} \quad (7)$$

where the Coulomb gauge is applied. Here, $m_0$ is the free electron mass, $\psi(\vec{r},t)$ is the electron wave function, $\vec{r}$ is the displacement vector, $t$ is time, $e$ is the electron charge, $\vec{A}(\vec{r},t)$ and $\varphi(\vec{r},t)$ are the electromagnetic vector and scalar potentials, respectively. Very often, the scalar potential $\varphi(\vec{r},t)$ is neglected. Additionally, when interaction of high energy electrons with optical modes of nanostructures and lasers are considered, it is convenient to use perturbation algorithms, such as adiabatic approximations [211]. Within adiabatic approximation one considers the wavefunction as $\psi(\vec{r},t) = \psi_0(\vec{r},t)\exp(iS(\vec{r},t)/\hbar)$, where the insertion of this ansatz in eq. (7) leads to a coupled system of equations for the unknowns $\psi_0(\vec{r},t)$ and $S(\vec{r},t)$, which is further simplified by considering the first orders in $\hbar$. Additionally, in many cases a major simplification is further applied, where the spatio-temporal changes in the amplitude is also neglected. The solutions to the electron wavefunction in this case are called Wolkow sates [212]. Using Wolkow assumption, all the changes in the electron wavefunction during its interaction with the electromagnetic field are accumulated in the phase of the electron as

$$\psi(\vec{r},t) = \frac{1}{(2\pi)^{\frac{3}{2}}}\exp\left( i\hbar\vec{k}_e\cdot\vec{r} - i\frac{\hbar^2 k_e^2}{2m_0}t \right) \times$$
$$\exp\left( -i\int_0^t d\tau \left( \frac{e\hbar\vec{k}_e\cdot\vec{A}(\tau)}{m_0} + \frac{e^2}{2m_0}\vec{A}^2(\tau) \right) \right) \quad (8)$$

Where $k_e = m_0\vec{V}_e/\hbar$ is the electron wave vector, and $\vec{V}_e$ is the initial velocity of the electron. This approximation has a wide range of applications in understanding the electron-photon interactions in photon-induced near-field electron microscopy [29], and in general in laser-assisted dynamics [213].

However, despite all the success of the adiabatic approximations in understanding the behavior of electrons interacting with magnetic domains and strong laser fields, there exists a range of phenomena where a full-wave analysis should be accounted. These phenomena include interaction of slow electrons with structures and fields in low-energy electron-diffraction [214, 215] and in general whenever a strong recoil and modulation of the amplitude of the electron wave function is expected, more accurate numerical methods are to be considered. Self-consistent numerical methods are examples of such techniques.

Self-consistent numerical methods are routinely applied to model numerous physical processes including (i) coupling of quantum emitters to dielectric and plasmonic nanoparticles within the empirical Maxwell–Bloch framework [216, 217], (ii) charge carrier mobility induced by the electromagnetic radiation in nanostructures and free-space within the Maxwell–Schrödinger framework [121, 218], and

(iii) charged-particle trajectories in free electron lasers and accelerators based on Maxwell–Lorentz combined systems of equations [219, 220]. Here, we focus on the interaction of free electrons with electromagnetic radiation to be exploited in accelerators and electron microscopes, and their modelling within the self-consistent Maxwell–Lorentz and Maxwell–Schrödinger frameworks.

### 6.1. Maxwell–Lorentz framework

Perhaps one of the most used numerical techniques is the particle-in-cell (PIC) method. This method is applied to a multitude of physical problems from plasma physics to radiation sources such as free electron lasers and accelerators. Moreover, PIC initiated the development in the field of self-consistent analysis, and historically rooted from the self-consistent calculations of Buneman [221] and Dawson [222] in the 1950s. In PIC the dynamics of the particles interacting with the radiation is computed in the phase space at continuous particle locations and is then projected on the grids associated with the discretized simulation domain for the field components. The simulation is called self-consistent mainly because the source terms, which appear in the Maxwell's equations because of the charged particle motions, are accumulated at each time step from the particle domain to the field-solver. The particle positions and dynamics are calculated using the relativistic Newton–Lorentz system of equations, whereas in a finite-difference form either the Boris scheme [223] or the Vay scheme [224] are adopted.

In the absence of external electromagnetic radiation, and especially when single-particle trajectories are considered beyond the non-recoil approximation, special care should be exerted when mapping the particle positions from the continuous to the discretized domains. This problem is at the heart of the problem of self-inertia and radiation resistance for electrons in classical electrodynamics [225]. Nevertheless, the mapping of a singular particle to the simulation domain is usually performed using interpolation schemes, such as nearest grid points, linear, and quadratic methods [219]. A more suitable method however, is to employ a spherically symmetric electron toy model without a strict radial cut-off, such as a Gaussian electron model [115, 150].

### 6.2. Maxwell–Schrödinger framework

Self-consistent analysis has recently been applied to investigate quantum-mechanical effects in modelling charge carrier mobility in low-dimensional quantum devices such as carbon nanotubes [216, 218, 226, 227]. In this context, a numerical self-consistent toolbox has been developed based on the combined system of Maxwell and Schrödinger equations, in order to study the dynamics of free-space single-electron wave packets interacting with nanostructures and light. While electron bunching has been exclusively considered within the context of classical framework (Lorentz equations) and many-particle dynamics, using the mentioned numerical toolbox it has been shown that even a single electron wave packet interacting with optical gratings and laser fields can undergo a bunching effect as well (figure 14(a)) [121]. Furthermore, It has been analytically demonstrated by Gover and Pan that the stimulated mechanism of radiation can have a direct influence on the shape of the electron wave packet up to a certain length of the electron drift [116]. Moreover, such a platform, which includes gratings of nanostructures and coherent radiation sources, can be used to shape the electron wave function into a desired form (figure 14(b)).

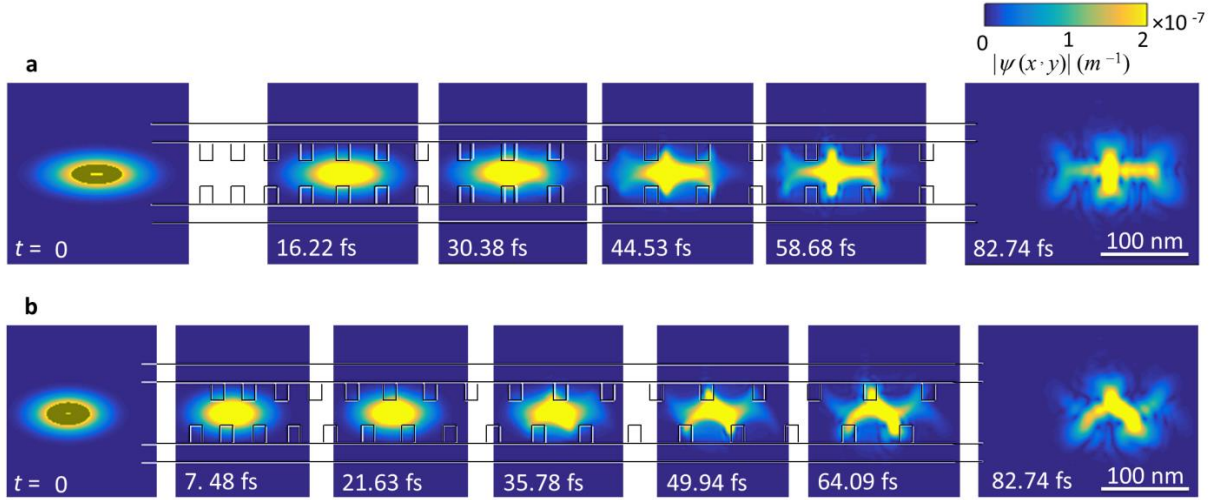

**Figure 14.** (a) Coherent interaction of a single electron wave packet with a chirped silicon grating, where the grating is excited symmetrically from both sides with a pulsed laser excitation at the carrier wavelength of 830 nm. Adapted from [121]. (b) A planar chirality of the wave function is achievable by electron interaction with an asymmetric grating.

## 7. Conclusion and Outlook

Overall, electron beams interacting with nanostructures cover a vast variety of mechanisms of interaction, and constitutes unparalleled probes for the optical density of states in nanostructures. However, carefully engineered nanostructure may be used also for manipulating the free-space states of the electron beam, in order to either shape it or to create new sources of tunable coherent light. Metamaterials might play an important role here, but still are to be further explored. In particular, the dominant contribution of bulk plasmons to the electron-induced radiation might render epsilon-near-zero materials into interesting candidates for the design of electron-driven photon sources [228]. Additionally, such radiation sources might be directly implemented into electron microscopes to provide new characterization techniques which involve both photon and electron beams. Ultrafast coherent control of physical and chemical processes within time-energy phase space and at a single nanostructure level can be achieved with concomitant utilization of electron and photon beams, demanding that the mutual coherent between them become achievable at the attosecond level. Within this context, theoretical and numerical methods play a vital role to further understand the electron–light–material interaction especially beyond the nonrecoil approximation and to capture the underlying dynamics. Finally, self-consistent numerical methods are required to take into account the self-field of the electron during the course of interaction.

**Appendix: Cherenkov radiation in anisotropic and permeable materials**

We assume an electron which is propagating inside a material at the velocity $V$ along the $z$-axis. The material is allowed to be anisotropic with the permittivity tensor:

$$\hat{\varepsilon}_r = \begin{bmatrix} \varepsilon_{rxx} & 0 & 0 \\ 0 & \varepsilon_{ryy} & 0 \\ 0 & 0 & \varepsilon_{rzz} \end{bmatrix}$$

and also permeable with the permeability $\mu_r$. We use the time-harmonic magnetic vector ($\vec{A}(\vec{r},\omega)$) and scalar ($\varphi(\vec{r},\omega)$) potentials, which are related to the field components as $\vec{B} = \vec{\nabla} \times \vec{A}$ and $\vec{E}(\vec{r},\omega) = -i\omega\vec{A}(\vec{r},\omega) - \vec{\nabla}\varphi(\vec{r},\omega)$ [104].

Using a convenient gauge theorem and the Maxwell equations, it is quite straight forward to construct a Helmholtz equation for $\vec{A}(\vec{r},\omega)$ as $\nabla^2 A_\alpha(\vec{r},\omega) + \varepsilon_{r\alpha\alpha}\mu_r k_0^2 A_\alpha = -\mu_0\mu_r \vec{J}(\vec{r},\omega)$ where $\alpha \in (x,y,z)$, and $\vec{J}(\vec{r},\omega)$ is the current density distribution associated with the swift electron and given by $J_z(\vec{r},\omega) = \int_{-\infty}^{+\infty} J_z(\vec{r},t) e^{i\omega t} dt = -eV \int_{-\infty}^{+\infty} \delta(x)\delta(y)\delta(z-Vt) e^{i\omega t} dt = \frac{-e}{4\pi^2} e^{i\frac{\omega}{V}z} \int_{-\infty}^{+\infty}\int_{-\infty}^{+\infty} e^{-ik_y y} e^{-ik_x x} dk_x dk_y$. For an electron moving in a bulk medium, the whole solution may be constructed with only one component of the vector potential as:

$$A_z^{inc}(\vec{r},\omega) = \int_{-\infty}^{+\infty} \frac{q\mu_0\mu_r}{2\pi} \frac{e^{i\frac{\omega}{V}z}}{\varepsilon_{rz}\mu_r k_0^2 - k_x^2 - k_y^2 - (\omega/V)^2} e^{jk_x x} e^{jk_y y} dk_x dk_y$$

which can be then used to calculate the field components, the EEL probability spectrum and the radiated power. However, for including material boundaries and interfaces, it is not sufficient to consider only one component of the vector potential as $\vec{A} = (0, 0, A_z)$, rather a more sophisticated choice of two components will be necessary [65].

**References**

[1] Erni R, Rossell M D, Kisielowski C and Dahmen U 2009 Atomic-resolution imaging with a sub-50-pm electron probe *Phys. Rev. Lett.* **102**
[2] de Abajo F J G 2010 Optical excitations in electron microscopy *Rev. Mod. Phys.* **82** 209-75
[3] Ritchie R H and Howie A 1988 Inelastic-scattering probabilities in scanning-transmission electron-microscopy *Philos. Mag. A* **58** 753-67
[4] de Abajo F J G and Kociak M 2008 Probing the photonic local density of states with electron energy loss spectroscopy *Phys. Rev. Lett.* **100** 106804-1_4
[5] Ogut B, Talebi N, Vogelgesang R, Sigle W and van Aken P A 2012 Toroidal plasmonic eigenmodes in oligomer nanocavities for the visible *Nano Lett.* **12** 5239-44
[6] Schmidt F P, Ditlbacher H, Hohenester U, Hohenau A, Hofer F and Krenn J R 2014 Universal dispersion of surface plasmons in flat nanostructures *Nat. Commun.* **5** 3604
[7] Barrow S J, Rossouw D, Funston A M, Botton G A and Mulvaney P 2014 Mapping bright and dark modes in gold nanoparticle chains using electron energy loss spectroscopy. *Nano Lett.* **14** 3799-808
[8] Schmidt F P, Ditlbacher H, Hohenester U, Hohenau A, Hofer F and Krenn J R 2012 Dark plasmonic breathing modes in silver nanodisks *Nano Lett.* **12** 5780-3

[9] Nelayah J, Kociak M, Stephan O, de Abajo F J G, Tence M, Henrard L, Taverna D, Pastoriza-Santos I, Liz-Marzan L M and Colliex C 2007 Mapping surface plasmons on a single metallic nanoparticle *Nat. Phys.* **3** 348-53
[10] DeJarnette D and Roper D K 2014 Electron energy loss spectroscopy of gold nanoparticles on graphene *J. Appl. Phys.* **116** 054313
[11] Nicoletti O, Wubs M, Mortensen N A, Sigle W, van Aken P A and Midgley P A 2011 Surface plasmon modes of a single silver nanorod: an electron energy loss study *Opt. Express* **19** 15371-9
[12] Boudarham G and Kociak M 2012 Modal decompositions of the local electromagnetic density of states and spatially resolved electron energy loss probability in terms of geometric modes *Phys. Rev. B* **85** 245447
[13] Ogut B, Vogelgesang R, Sigle W, Talebi N, Koch C T and van Aken P A 2011 Hybridized metal slit eigenmodes as an illustration of babinet's principle *ACS Nano* **5** 6701-6
[14] Walther R, Fritz S, Mueller E, Schneider R, Gerthsen D, Sigle W, Maniv T, Cohen H, Matyssek C and Busch K 2016 Coupling of surface-plasmon-polariton-hybridized cavity modes between submicron slits in a thin gold film *ACS Photonics* **3** 836-43
[15] Salomon A, Prior Y, Fedoruk M, Feldmann J, Kolkowski R and Zyss J 2014 Plasmonic coupling between metallic nanocavities *J. Opt. (Bristol, U. K.)* **16** 114012
[16] Talebi N, Ogut B, Sigle W, Vogelgesang R and van Aken P A 2014 On the symmetry and topology of plasmonic eigenmodes in heptamer and hexamer nanocavities *Appl. Phys. A: Mater. Sci. Process* **116** 947-54
[17] Gu L, Sigle W, Koch C T, Ogut B, van Aken P A, Talebi N, Vogelgesang R, Mu J L, Wen X G and Mao J 2011 Resonant wedge-plasmon modes in single-crystalline gold nanoplatelets *Phys. Rev. B* **83** 195433
[18] Xu X B, Yi Z, Li X B, Wang Y Y, Liu J P, Luo J S, Luo B C, Yi Y G and Tang Y J 2013 Tunable nanoscale confinement of energy and resonant edge effect in triangular gold nanoprisms *J. Phys. Chem. C* **117** 17748-56
[19] Bellido E P, Manjavacas A, Zhang Y, Cao Y, Nordlander P and Botton G A 2016 Electron energy-loss spectroscopy of multipolar edge and cavity modes in silver nanosquares *ACS Photonics* **3** 428-33
[20] Yamamoto N, de Abajo F J G and Myroshnychenko V 2015 Interference of surface plasmons and Smith-Purcell emission probed by angle-resolved cathodoluminescence spectroscopy *Phys. Rev. B* **91** 125144
[21] Talebi N, Sigle W, Vogelgesang R, Esmann M, Becker S F, Lienau C and van Aken P A 2015 Excitation of mesoscopic plasmonic tapers by relativistic electrons: phase matching versus eigenmode resonances *ACS Nano* **9** 7641-8
[22] Guo S R, Talebi N, Sigle W, Vogelgesang R, Richter G, Esmann M, Becker S F, Lienau C and van Aken P A 2016 Reflection and phase matching in plasmonic gold tapers *Nano Lett.* **16** 6137-44
[23] Glenn D R, Zhang H, Kasthuri N, Schalek R, Lo P K, Trifonov A S, Park H, Lichtman J W and Walsworth R L 2012 Correlative light and electron microscopy using cathodoluminescence from nanoparticles with distinguishable colours *Sci. Rep.* **2** 865
[24] Kociak M, Stephan O, Gloter A, Zagonel L F, Tizei L H G, Tence M, March K, Blazit J D, Mahfoud Z, Losquin A, Meuret S and Colliex C 2014 Seeing and measuring in colours: Electron microscopy and spectroscopies applied to nano-optics *C. R. Phys.* **15** 158-75
[25] Coenen T, den Hoedt S V and Polman A 2016 A new cathodoluminescence system for nanoscale optics, materials science, and geology *Microscopy Today* **24** 12-9
[26] de Abajo F J G, Sapienza R, Noginov M, Benz F, Baumberg J, Maier S, Graham D, Aizpurua J, Ebbesen T, Pinchuk A, Khurgin J, Matczyszyn K, Hugall J T, van Hulst N, Dawson P, Roberts C, Nielsen M, Bursi L, Flatte M, Yi J, Hess O, Engheta N, Brongersma M, Podolskiy V, Shalaev V,

Narimanov E and Zayats A 2015 Plasmonic and new plasmonic materials: general discussion *Faraday Discuss* **178** 123-49

[27] Liu Y M and Zhang X 2011 Metamaterials: a new frontier of science and technology *Chem. Soc. Rev.* **40** 2494-507

[28] Engheta N and Ziolkowski R W 2005 A positive future for double-negative metamaterials *IEEE Trans. Microwave Theory Tech.* **53** 1535-56

[29] Park S T, Lin M M and Zewail A H 2010 Photon-induced near-field electron microscopy (PINEM): theoretical and experimental *New J. Phys.* **12** 123028

[30] Barwick B and Zewail A H 2015 Photonics and plasmonics in 4D ultrafast electron microscopy *ACS Photonics* **2** 1391-402

[31] Pitarke J M, Silkin V M, Chulkov E V and Echenique P M 2007 Theory of surface plasmons and surface-plasmon polaritons *Rep. Prog. Phys.* **70** 1

[32] Ozbay E 2006 Plasmonics: Merging photonics and electronics at nanoscale dimensions *Science* **311** 189-93

[33] Ritchie R H 1957 Plasma Losses by Fast Electrons in Thin Films *Phys. Rev.* **106** 874-81

[34] Powell C J and Swan J B 1959 Origin of the Characteristic Electron Energy Losses in Aluminum *Phys. Rev.* **115** 869-75

[35] Powell C J and Swan J B 1959 Origin of the Characteristic Electron Energy Losses in Magnesium *Phys. Rev.* **116** 81-3

[36] Basov D N, Fogler M M and de Abajo F J G 2016 Polaritons in van der Waals materials *Science* **354** aag1992-8

[37] Gramotnev D K and Bozhevolnyi S I 2010 Plasmonics beyond the diffraction limit *Nat. Photon.* **4** 83-91

[38] Grubisic A, Ringe E, Cobley C M, Xia Y N, Marks L D, Van Duyne R P and Nesbitt D J 2012 Plasmonic Near-Electric Field Enhancement Effects in Ultrafast Photoelectron Emission: Correlated Spatial and Laser Polarization Microscopy Studies of Individual Ag Nanocubes *Nano Lett.* **12** 4823-9

[39] Kauranen M and Zayats A V 2012 Nonlinear plasmonics *Nat. Photon.* **6** 737-48

[40] Schuller J A, Barnard E S, Cai W S, Jun Y C, White J S and Brongersma M L 2010 Plasmonics for extreme light concentration and manipulation *Nat. Mater.* **9** 193-204

[41] Chung T, Lee S Y, Song E Y, Chun H and Lee B 2011 Plasmonic Nanostructures for Nano-Scale Bio-Sensing *Sensors (Basel)* **11** 10907-29

[42] Krasnok A E, Maksymov I S, Denisyuk A I, Belov P A, Miroshnichenko A E, Simovski C R and Kivshar Y S 2013 Optical nanoantennas *Physics-Uspekhi* **56** 539-64

[43] Stokes J L, Sarua A, Pugh J R, Dorh N, Munns J W, Bassindale P G, Ahmad N, Orr-Ewing A J and Cryan M J 2015 Purcell enhancement and focusing effects in plasmonic nanoantenna arrays *J. Opt. Soc. Am. B* **32** 2158-63

[44] Pendry J B, Martin-Moreno L and Garcia-Vidal F J 2004 Mimicking surface plasmons with structured surfaces *Science* **305** 847-8

[45] Talebi N and Shahabadi M 2010 Spoof surface plasmons propagating along a periodically corrugated coaxial waveguide *J. Phys. D Appl. Phys.* **43** 135302

[46] Monticone F and Alu A 2014 Metamaterials and plasmonics: From nanoparticles to nanoantenna arrays, metasurfaces, and metamaterials *Chinese Phys. B* **23** 047809

[47] Yao K and Liu Y M 2014 Plasmonic metamaterials *Nanotechnol. Rev.* **3** 177-210

[48] Shalaev V M 2007 Optical negative-index metamaterials *Nat. Photon.* **1** 41-8

[49] Soukoulis C M and Wegener M 2011 Past achievements and future challenges in the development of three-dimensional photonic metamaterials *Nat. Photon.* **5** 523-30

[50] Kapitza P L and Dirac P A M 1933 The reflection of electrons from standing light waves *Math. Proc. Cambridge Philos. Soc.* **29** 297-300
[51] Batelaan H 2007 Colloquium: Illuminating the Kapitza-Dirac effect with electron matter optics *Rev. Mod. Phys.* **79** 929-41
[52] Wu S J, Wang Y J, Diot Q and Prentiss M 2005 Splitting matter waves using an optimized standing-wave light-pulse sequence *Phys. Rev. A* **71** 043602
[53] Rasel E M, Oberthaler M K, Batelaan H, Schmiedmayer J and Zeilinger A 1995 Atom wave interferometry with diffraction gratings of light *Phys. Rev. Lett.* **75** 2633-7
[54] Hayrapetyan A G, Grigoryan K K, Gotte J B and Petrosyan R G 2015 Kapitza-Dirac effect with traveling waves *New J. Phys.* **17** 082002
[55] Freimund D L, Aflatooni K and Batelaan H 2001 Observation of the Kapitza-Dirac effect *Nature* **413** 142-3
[56] Friedman A, Gover A, Kurizki G, Ruschin S and Yariv A 1988 Spontaneous and stimulated-emission from quasifree electrons *Rev. Mod. Phys.* **60** 471-535
[57] Gover A and Sprangle P 1981 A unified theory of magnetic Bremsstrahlung, electrostatic Bremsstrahlung, Compton-Raman scattering, and Cerenkov-Smith-Purcell free-electron lasers *IEEE J. Quantum Electron.* **17** 1196-215
[58] Bucksbaum P, Moller T and Ueda K 2013 Frontiers of free-electron laser science *J. Phys. B: At. Mol. Opt. Phys.* **46** 160201
[59] Falcone R, Dunne M, Chapman H, Yabashi M and Ueda K 2016 Frontiers of free-electron laser science II *J. Phys. B: At. Mol. Opt. Phys.* **49** 180201
[60] Huang Z R and Kim K J 2007 Review of x-ray free-electron laser theory *Phys. Rev. ST Accel. Beams* **10** 034801
[61] McNeil B W J and Thompson N R 2010 X-ray free-electron lasers *Nat. Photonics* **4** 814-21
[62] Melrose D B, Ronnmark K G and Hewitt R G 1982 Terrestrial kilometric radiation - the cyclotron Ttheory *J. Geophys. Res.* **87** 5140-50
[63] Asner D M, Bradley R F, de Viveiros L, Doe P J, Fernandes J L, Fertl M, Finn E C, Formaggio J A, Furse D, Jones A M, Kofron J N, LaRoque B H, Leber M, McBride E L, Miller M L, Mohanmurthy P, Monreal B, Oblath N S, Robertson R G H, Rosenberg L J, Rybka G, Rysewyk D, Sternberg M G, Tedeschi J R, Thummler T, VanDevender B A and Woods N L 2015 Single-electron detection and spectroscopy via relativistic cyclotron radiation *Phys. Rev. Lett.* **114** 162501
[64] Ginzburg V L 1996 Radiation of uniformly moving sources (Vavilov-Cherenkov effect, transition radiation, and other phenomena) *Phys.-Usp.* **39** 973
[65] Garciamolina R, Grasmarti A, Howie A and Ritchie R H 1985 Retardation effects in the interaction of charged-particle beams with bounded condensed media *J. Phys. C: Solid State Phys.* **18** 5335
[66] Kröger E 1968 Calculations of the energy losses of fast electrons in thin foils with retardation *Z. Physik* **216** 115-35
[67] Cherenkov P A 1938 The spectrum of visible radiation produced by fast electrons *C. R. (Dokl.) Acad. Sci. URSS* **20** 651-5
[68] Cherenkov P A 1938 Absolute output of radiation caused by electrons moving within a medium with super-light velocity *C. R. (Dokl.) Acad. Sci. URSS* **21** 116-21
[69] Cherenkov P A 1938 Spatial distribution of visible radiation produced by fast electrons *C. R. (Dokl.) Acad. Sci. URSS* **21** 319-21
[70] Frank I and Tamm I 1937 Coherent visible radiation of fast electrons passing through matter *Cr Acad Sci Urss* **14** 109-14
[71] Li W, Yu C-X and Liu S-B 2009 Quantum theory of Cherenkov radiation in an anisotropic absorbing media. *Proc. SPIE* **7501** 750108--8

[72]   Kaminer I, Mutzafi M, Levy A, Harari G, Herzig Sheinfux H, Skirlo S, Nemirovsky J, Joannopoulos J D, Segev M and Soljačić M 2016 Quantum cherenkov radiation: spectral cutoffs and the role of spin and orbital angular momentum *Phys. Rev. X* **6** 011006

[73]   Matloob R and Ghaffari A 2004 Cerenkov radiation in a causal permeable medium *Phys. Rev. A* **70** 052116

[74]   Chefranov S G 2004 Relativistic generalization of the Landau criterion as a new foundation of the Vavilov-Cherenkov radiation theory *Phys. Rev. Lett.* **93** 269902

[75]   Tanha K, Pashazadeh A M and Pogue B W 2015 Review of biomedical Cerenkov luminescence imaging applications *Biomed Opt. Express* **6** 3053-65

[76]   Buchanan M 2007 Thesis: Minkowski, Abraham and the photon momentum *Nat Phys* **3** 73-

[77]   Festenberg C v 1969 Energy loss measurements on III–V compounds *Zeitschrift für Physik* **227** 453-81

[78]   Chen C H, Silcox J and Vincent R 1975 Electron-Energy Losses in Silicon - Bulk and Surface Plasmons and Cerenkov Radiation *Phys. Rev. B* **12** 64-71

[79]   Stoger-Pollach M, Franco H, Schattschneider P, Lazar S, Schaffer B, Grogger W and Zandbergen H W 2006 Cerenkov losses: A limit for bandgap determination and Kramers-Kronig analysis *Micron* **37** 396-402

[80]   Schattschneider P 1986 *Fundamentals of inelastic electron scattering* (Wien: Springer-Verlag)

[81]   Coenen T, Vesseur E J R and Polman A 2011 Angle-resolved cathodoluminescence spectroscopy *Appl. Phys. Lett.* **99** 143103

[82]   Coenen T, Vesseur E J R, Polman A and Koenderink A F 2011 Directional Emission from Plasmonic Yagi-Uda Antennas Probed by Angle-Resolved Cathodoluminescence Spectroscopy *Nano Lett.* **11** 3779-84

[83]   Talebi N, Ozsoy-Keskinbora C, Benia H M, Kern K, Koch C T and van Aken P A 2016 Wedge Dyakonov Waves and Dyakonov Plasmons in Topological Insulator Bi2Se3 Probed by Electron Beams *Acs Nano* **10** 6988-94

[84]   Esslinger M, Vogelgesang R, Talebi N, Khunsin W, Gehring P, de Zuani S, Gompf B and Kern K 2014 Tetradymites as Natural Hyperbolic Materials for the Near-Infrared to Visible *Acs Photonics* **1** 1285-9

[85]   Smith D R and Kroll N 2000 Negative refractive index in left-handed materials *Phys. Rev. Lett.* **85** 2933-6

[86]   Veselago V G 1968 The electrodynamics of substances with simultaneously negative values of ε and μ *Sov. Phys. Usp.* **10** 509

[68]   Galyamin S N, Tyukhtin A V, Kanareykin A and Schoessow P 2009 Reversed Cherenkov-transition radiation by a charge crossing a left-handed medium boundary *Phys. Rev. Lett.* **103** 194802

[69]   Grbic A and Eleftheriades G V 2002 Experimental verification of backward-wave radiation from a negative refractive index metamaterial *J. Appl. Phys.* **92** 5930-5

[89]   Liu F, Xiao L, Ye Y, Wang M X, Cui K Y, Feng X, Zhang W and Huang Y D 2017 Integrated Cherenkov radiation emitter eliminating the electron velocity threshold *Nat. Photon.* **11** 289

[90]   Ginsburg V and Frank I 1946 Radiation of a uniformly moving electron due to its transition from one medium into another *Zh.Eksp.Teor.Fiz.* **16** 15-28

[91]   Ginzburg V L 1982 Transition radiation and transition scattering *Phys. Scr.* **1982** 182

[92]   Ginzburg V L and Tsytovich V N 1979 Several problems of the theory of transition radiation and transition scattering *Phys. Rep.* **49** 1-89

[93]   de Abajo F J G, Rivacoba A, Zabala N and Yamamoto N 2004 Boundary effects in Cherenkov radiation *Phys. Rev. B* **69** 155420

[94]   Happek U, Sievers A J and Blum E B 1991 Observation of coherent transition radiation *Phys. Rev. Lett.* **67** 2962-5

[95] Glinec Y, Faure J, Norlin A, Pukhov A and Malka V 2007 Observation of fine structures in laser-driven electron beams using coherent transition radiation *Phys. Rev. Lett.* **98** 194801
[96] Lumpkin A H, Dejus R, Berg W J, Borland M, Chae Y C, Moog E, Sereno N S and Yang B X 2001 First observation of z-dependent electron-beam microbunching using coherent transition radiation *Phys. Rev. Lett.* **86** 79-82
[97] Tremaine A, Rosenzweig J B, Anderson S, Frigola P, Hogan M, Murokh A, Pellegrini C, Nguyen D C and Sheffield R L 1998 Observation of self-amplified spontaneous-emission-induced electron-beam microbunching using coherent transition radiation *Phys. Rev. Lett.* **81** 5816-9
[98] Shibata Y, Takahashi T, Kanai T, Ishi K, Ikezawa M, Ohkuma J, Okuda S and Okada T 1994 Diagnostics of an electron-beam of a linear-accelerator using coherent transition radiation *Phys. Rev. E* **50** 1479-84
[99] Wu Z R, Fisher A S, Goodfellow J, Fuchs M, Daranciang D, Hogan M, Loos H and Lindenberg A 2013 Intense terahertz pulses from SLAC electron beams using coherent transition radiation *Rev. Sci. Instrum.* **84** 022701
[100] Ding W J and Sheng Z M 2016 Sub GV/cm terahertz radiation from relativistic laser-solid interactions via coherent transition radiation *Phys. Rev. E* **93** 063204
[101] Egerton R F 2009 Electron energy-loss spectroscopy in the TEM *Rep. Prog. Phys.* **72** 016502
[102] Wartski L, Roland S, Lasalle J, Bolore M and Filippi G 1975 Interference phenomenon in optical transition radiation and its application to particle beam diagnostics and multiple-scattering measurements *J. Appl. Phys.* **46** 3644-53
[103] Vincent R and Silcox J 1973 Dispersion of radiative surface plasmons in aluminum films by electron-scattering *Phys. Rev. Lett.* **31** 1487-90
[104] Talebi N 2016 Optical modes in slab waveguides with magnetoelectric effect *J. Opt. (Bristol, U. K.)* **18** 055607
[105] Jacob Z, Kim J Y, Naik G V, Boltasseva A, Narimanov E E and Shalaev V M 2010 Engineering photonic density of states using metamaterials *Appl. Phys. B* **100** 215-8
[106] Cortes C L, Newman W, Molesky S and Jacob Z 2012 Quantum nanophotonics using hyperbolic metamaterials *J. Opt. (Bristol, U. K.)* **14** 063001
[107] Schachter L and Ron A 1989 Smith-Purcell Free-Electron Laser *Phys. Rev. A* **40** 876-96
[108] Wang M H, Xiao X G, Chen J Y and Wei Y Y 2005 Study on a novel Smith-Purcell free-electron laser *Phys. Lett. A* **345** 423-7
[109] Brenny B J M, Polman A and García de Abajo F J 2016 Femtosecond plasmon and photon wave packets excited by a high-energy electron on a metal or dielectric surface *Phys. Rev. B* **94** 155412
[110] Novotny L 2010 Strong coupling, energy splitting, and level crossings: A classical perspective *Am. J. Phys.* **78** 1199-202
[111] Rodriguez S R K 2016 Classical and quantum distinctions between weak and strong coupling *Eur. J. Phys.* **37** 025802
[112] Smith S J and Purcell E M 1953 Visible light from localized surface charges moving across a grating *Phys. Rev.* **92** 1069
[113] Goldsmith P and Jelley J V 1959 Optical transition radiation from protons entering metal surfaces *Philos Mag* **4** 836-44
[114] de Abajo F J G 2000 Smith-Purcell radiation emission in aligned nanoparticles *Phys. Rev. E* **61** 5743-52
[115] Talebi N 2014 A directional, ultrafast and integrated few-photon source utilizing the interaction of electron beams and plasmonic nanoantennas *New J. Phys.* **16**
[116] Gover A and Pan Y 2017 Stimulated radiation interaction of a single electron quantum wavepacket. In: *ArXiv e-prints*,

[117] Andrews H L, Boulware C H, Brau C A and Jarvis J D 2005 Superradiant emission of Smith-Purcell radiation *Phys. Rev. ST. Accel. Beams* **8** 110702
[118] Dicke R H 1954 Coherence in spontaneous radiation processes *Phys. Rev.* **93** 99-110
[119] Phillips R M 1988 History of the Uubitron *Nucl. Instrum. Methods Phys. Res., Sect. A* **272** 1-9
[120] Madey J M J 1971 Stimulated emission of bremsstrahlung in a periodic magnetic field *J. Appl. Phys.* **42** 1906
[121] Talebi N 2016 Schrödinger electrons interacting with optical gratings: A quantum mechanical study of inverse Smith Purcell effect *New J. Phys.* **18** 123006
[122] Korbly S E, Kesar A S, Sirigiri J R and Temkin R J 2005 Observation of frequency-locked coherent terahertz Smith-Purcell radiation *Phys. Rev. Lett* **94** 054803
[123] Li Y L, Sun Y E and Kim K J 2008 High-power beam-based coherently enhanced THz radiation source *Phys. Rev. Spec. Top.--Accel. Beams* **11** 080701
[124] Ozyuzer L, Koshelev A E, Kurter C, Gopalsami N, Li Q, Tachiki M, Kadowaki K, Yamamoto T, Minami H, Yamaguchi H, Tachiki T, Gray K E, Kwok W K and Welp U 2007 Emission of coherent THz radiation from superconductors *Science* **318** 1291-3
[125] Meng X Z 2013 Smith-Purcell free electron laser based on a multilayer metal-dielectric stack *Optik* **124** 3162-4
[126] Meng X Z, Wang M H and Ren Z M 2011 Smith-Purcell free electron laser based on the semi-elliptical resonator *Chinese Phys. B* **20**
[127] Prokop C, Piot P, Lin M C and Stoltz P 2010 Numerical modeling of a table-top tunable Smith-Purcell terahertz free-electron laser operating in the super-radiant regime *Appl. Phys. Lett.* **96** 151502
[128] Wang M H, Liu P K, Ge G Y and Dong R X 2007 Free electron laser based on the Smith-Purcell radiation *Opt. Laser Technol.* **39** 1254-7
[129] Liu W X, Yang Z Q, Liang Z, Li D and Imasaki K 2007 Two-stream Smith-Purcell free-electron laser *Nucl. Instrum. Methods Phys. Res., Sect. A* **570** 171-5
[130] Kumar V and Kim K J 2006 Analysis of Smith-Purcell free-electron lasers *Phys. Rev. E* **73** 026501
[131] Chen J W, Fu S F and Zhang D K 1985 Smith-Purcell free-electron laser with variable grating period *Chinese. Phys.* **5** 194-8
[132] Wachtel J M 1979 Free-Electron Lasers Using the Smith-Purcell Effect *J. Appl. Phys.* **50** 49-56
[133] Leavitt R P, Wortman D E and Morrison C A 1979 Orotron - free-electron laser using the smith-purcell effect *Appl. Phys. Lett.* **35** 363-5
[134] Gover A and Livni Z 1978 Operation regimes of Cerenkov - Smith-Purcell free-electron lasers and TW amplifiers *Opt. Commun.* **26** 375-80
[135] Kaminer I, Kooi S E, Shiloh R, Zhen B, Shen Y, Lopez J J, Remez R, Skirlo S A, Yang Y, Joannopoulos J D, Arie A and Soljacic M 2017 Spectrally and Spatially Resolved Smith-Purcell Radiation in Plasmonic Crystals with Short-Range Disorder *Physical Review X* **7** 011003
[136] Bratman V L, Ginzburg N S and Petelin M I 1979 Common Properties of Free-Electron Lasers *Opt. Commun.* **30** 409-12
[137] Woods K J, Walsh J E, Stoner R E, Kirk H G and Fernow R C 1995 Forward directed Smith-Purcell radiation from relativistic electrons *Phys. Rev. Lett.* **74** 3808-11
[138] Urata J, Goldstein M, Kimmitt M F, Naumov A, Platt C and Walsh J E 1998 Superradiant Smith-Purcell emission *Phys. Rev. Lett.* **80** 516-9
[139] Andrews H L, Boulware C H, Brau C A and Jarvis J D 2005 Dispersion and attenuation in a Smith-Purcell free electron laser *Phys. Rev. ST Accel. Beams* **8** 050703
[140] Bratman V L, Dumesh B S, Fedotov A E, Makhalov P B, Movshevich B Z and Rusin F S 2010 Terahertz orotrons and oromultipliers *IEEE Trans. Plasma Sci.* **38** 1466-71

[141] Liu W H, Lu Y L, Wang L and Jia Q K 2016 A multimode terahertz-orotron with the special Smith-Purcell radiation *Appl. Phys. Lett.* **108** 183510

[142] Kumar P, Bhasin L, Tripathi V K, Kumar A and Kumar M 2016 Smith-Purcell terahertz radiation from laser modulated electron beam over a metallic grating *Phys. Plasmas* **23** 093301

[143] Wang Z J, Yao K, Chen M, Chen H S and Liu Y M 2016 Manipulating Smith-Purcell Emission with Babinet Metasurfaces *Phys Rev Lett* **117** 157401

[144] Zhou Y C, Zhang Y X and Liu S G 2016 Electron-beam-driven enhanced terahertz coherent Smith-Purcell radiation within a cylindrical quasi-optical cavity *IEEE Trans. Terahertz Sci. Technol.* **6** 262-7

[145] Bendana X, Polman A and de Abajo F J G 2011 Single-photon generation by electron beams *Nano Lett.* **11** 5099-103

[146] Adamo G, MacDonald K F, Fu Y H, Wang C M, Tsai D P, de Abajo F J G and Zheludev N I 2009 Light well: a tunable free-electron light source on a chip *Phys. Rev. Lett.* **103** 113901

[147] Adamo G, MacDonald K F, Fu Y H, Tsai D P, de Abajo F J G and Zheludev N I 2010 Tuneable electron-beam-driven nanoscale light source *J. Opt. (Bristol, U. K.)* **12** 024012

[148] Adamo G, Ou J Y, So J K, Jenkins S D, De Angelis F, MacDonald K F, Di Fabrizio E, Ruostekoski J and Zheludev N I 2012 Electron-beam-driven collective-mode metamaterial light source *Phys. Rev. Lett.* **109** 217401

[149] Li G, Clarke B P, So J-K, MacDonald K F and Zheludev N I 2016 Holographic free-electron light source *Nature Communications* **7** 13705

[150] Talebi N 2016 Spectral interferometry with electron microscopes *Sci. Rep.* **6** 33874

[151] Tizei L H G and Kociak M 2013 Spatially resolved quantum nano-optics of single photons using an electron microscope *Phys. Rev. Lett.* **110** 153604

[152] Mahfoud Z, Dijksman A T, Javaux C, Bassoul P, Baudrion A L, Plain J, Dubertret B and Kociak M 2013 Cathodoluminescence in a scanning transmission electron microscope: A nanometer-scale counterpart of photoluminescence for the study of II-VI quantum dots *J. Phys. Chem. Lett.* **4** 4090-4

[153] Meuret S, Tizei L H G, Cazimajou T, Bourrellier R, Chang H C, Treussart F and Kociak M 2015 Photon bunching in cathodoluminescence *Phys. Rev. Lett.* **114** 197401

[154] Bourrellier R, Meuret S, Tararan A, Stephan O, Kociak M, Tizei L H G and Zobelli A 2016 Bright UV Single photon emission at point defects in h-BN *Nano Lett.* **16** 4317-21

[155] Talebi N, Sigle W, Vogelgesang R and van Aken P 2013 Numerical simulations of interference effects in photon-assisted electron energy-loss spectroscopy *New J. Phys.* **15**

[156] England R J, Noble R J, Bane K, Dowell D H, Ng C K, Spencer J E, Tantawi S, Wu Z R, Byer R L, Peralta E, Soong K, Chang C M, Montazeri B, Wolf S J, Cowan B, Dawson J, Gai W, Hommelhoff P, Huang Y C, Jing C, McGuinness C, Palmer R B, Naranjo B, Rosenzweig J, Travish G, Mizrahi A, Schachter L, Sears C, Werner G R and Yoder R B 2014 Dielectric laser accelerators *Rev. Mod. Phys.* **86** 1337-89

[157] Shimoda K 1962 Proposal for an Electron Accelerator Using an Optical Maser *Appl. Opt.* **1** 33-5

[158] Mizuno K, Pae J, Nozokido T and Furuya K 1987 Experimental-evidence of the inverse Smith-Purcell effect *Nature* **328** 45-7

[159] Esarey E, Sprangle P, Krall J and Ting A 1996 Overview of plasma-based accelerator concepts *IEEE Trans. Plasma Sci.* **24** 252-88

[160] Bae J, Shirai H, Nishida T, Nozokido T, Furuya K and Mizuno K 1992 Experimental verification of the theory on the inverse Smith–Purcell effect at a submillimeter wavelength *Appl. Phys. Lett.* **61** 870-2

[161] Saito N and Ogata A 2003 Plasmon linac: A laser wake-field accelerator based on a solid-state plasma *Phys. Plasmas* **10** 3358-62

[162] So J K, de Abajo F J G, MacDonald K F and Zheludev N I 2015 Amplification of the evanescent field of free electrons *ACS Photonics* **2** 1236-40

[163] Danisi A, Masi A and Losito R 2015 Numerical analysis of metamaterial insertions for mode damping in parasitic particle accelerator cavities *9th International Congress on Advanced Electromagnetic Materials in Microwaves and Optics (Metamaterials 2015)* 46-8

[164] Tanabe K 2008 Field enhancement around metal nanoparticles and nanoshells: A systematic investigation *J Phys. Chem. C* **112** 15721-8

[165] Bermel P, Byer R L, Colby E R, Cowan B M, Dawson J, England R J, Noble R J, Qi M H and Yoder R B 2014 Summary of the 2011 Dielectric laser accelerator workshop *Nucl. Instrum. Meth. A* **734** 51-9

[166] Mourou G, Brocklesby B, Tajima T and Limpert J 2013 The future is fibre accelerators *Nat. Photonics* **7** 258-61

[167] Wootton K P, Wu Z R, Cowan B M, Hanuka A, Makasyuk I V, Peralta E A, Soong K, Byer R L and England R J 2016 Demonstration of acceleration of relativistic electrons at a dielectric microstructure using femtosecond laser pulses *Opt. Lett.* **41** 2696-9

[168] McNeur J, Kozak M, Ehberger D, Schonenberger N, Tafel A, Li A and Hommelhoff P 2016 A miniaturized electron source based on dielectric laser accelerator operation at higher spatial harmonics and a nanotip photoemitter *J. Phys. B: At. Mol. Opt. Phys.* **49** 03400

[169] Breuer J, McNeur J and Hommelhoff P 2014 Dielectric laser acceleration of electrons in the vicinity of single and double grating structures-theory and simulations *J. Phys. B: At. Mol. Opt. Phys.* **47** 234004

[170] Breuer J, Graf R, Apolonski A and Hommelhoff P 2014 Dielectric laser acceleration of nonrelativistic electrons at a single fused silica grating structure: Experimental part *Phys. Rev. ST Accel Beams* **17** 021301

[171] Wu Z, England R J, Ng C-K, Cowan B, McGuinness C, Lee C, Qi M and Tantawi S 2014 Coupling power into accelerating mode of a three-dimensional silicon woodpile photonic band-gap waveguide *Phys. Rev. ST Accel Beams* **17** 081301

[172] Wei Y L, Xia G X, Smith J D A, Hanahoe K, Mete O, Jamison S P and Welsch C P 2015 Numerical study of a multi-stage dielectric laser-driven accelerator *Physcs. Proc.* **77** 50-7

[173] McNeur J, Kozak M, Schonenberger N, Li A, Tafel A and Hommelhoff P 2016 Laser-driven acceleration of subrelativistic electrons near a nanostructured dielectric grating: From acceleration via higher spatial harmonics to necessary elements of a dielectric accelerator *Nucl. Instrum. Meth. A* **829** 50-1

[174] Breuer J and Hommelhoff P 2013 Laser-based acceleration of nonrelativistic electrons at a dielectric structure *Phys. Rev. Lett.* **111** 134803

[175] Naranjo B, Valloni A, Putterman S and Rosenzweig J B 2012 Stable charged-particle acceleration and focusing in a laser accelerator using spatial harmonics *Phys. Rev. Lett.* **109** 164803

[176] Staude I, McGuinness C, Frolich A, Byer R L, Colby E and Wegener M 2012 Waveguides in three-dimensional photonic bandgap materials for particle-accelerator on a chip architectures *Opt. Express* **20** 5607-12

[177] Kofink W 1937 Über das magnetische und elektrische Moment des Elektrons nach der Diracschen Theorie *Ann. Phys. (Berlin, Ger.)* **422** 91-8

[178] Barnothy J 1948 On the intrinsic moment of the electron *Phys. Rev.* **74** 113

[179] Foley H M and Kusch P 1948 On the Intrinsic Moment of the Electron *Phys. Rev.* **73** 412

[180] De Rafael E 1971 The hydrogen hyperfine structure and inelastic electron proton scattering experiments *Phys. Lett. B* **37** 201-3

[181] Ohanian H C 1986 What Is Spin *Am. J. Phys.* **54** 500-5

[182] Harris J, Grillo V, Mafakheri E, Gazzadi G C, Frabboni S, Boyd R W and Karimi E 2015 Structured quantum waves *Nat. Phys.* **11** 629-34
[183] Rother A and Scheerschmidt K 2009 Relativistic effects in elastic scattering of electrons in TEM *Ultramicroscopy* **109** 154-60
[184] Shiloh R, Lereah Y, Lilach Y and Arie A 2014 Sculpturing the electron wave function using nanoscale phase masks *Ultramicroscopy* **144** 26-31
[185] Verbeeck J, Guzzinati G, Clark L, Juchtmans R, Van Boxem R, Tian H, Béché A, Lubk A and Van Tendeloo G 2014 Shaping electron beams for the generation of innovative measurements in the (S)TEM *C. R. Phys.* **15** 190-9
[186] Berry M V and Balazs N L 1979 Nonspreading wave packets *Am. J. Phys.* **47** 264-7
[187] Grillo V, Harris J, Gazzadi G C, Balboni R, Mafakheri E, Dennis M R, Frabboni S, Boyd R W and Karimi E 2016 Generation and application of bessel beams in electron microscopy *Ultramicroscopy* **166** 48-60
[188] Saitoh K, Hirakawa K, Nambu H, Tanaka N and Uchida M 2016 Generation of electron Bessel beams with nondiffractive spreading by a nanofabricated annular slit *J. Phys. Soc. Jpn.* **85**
[189] Voloch-Bloch N, Lereah Y, Lilach Y, Gover A and Arie A 2013 Generation of electron Airy beams *Nature* **494** 331-5
[190] Guzzinati G, Béché A, Lourenço-Martins H, Martin J, Kociak M and Verbeeck J 2017 Probing the symmetry of the potential of localized surface plasmon resonances with phase-shaped electron beams *Nature Communications* **8** 14999
[191] Nakamura Y, Pashkin Y A and Tsai J S 1999 Coherent control of macroscopic quantum states in a single-Cooper-pair box *Nature* **398** 786-8
[192] Rabitz H, de Vivie-Riedle R, Motzkus M and Kompa K 2000 Chemistry - Whither the future of controlling quantum phenomena? *Science* **288** 824-8
[193] Handali J, Shakya P and Barwick B 2015 Creating electron vortex beams with light *Opt Express* **23** 5236-43
[194] Bliokh K Y, Bliokh Y P, Savel'ev S and Nori F 2007 Semiclassical dynamics of electron wave packet states with phase vortices *Phys Rev Lett* **99** 190404
[195] Uchida M and Tonomura A 2010 Generation of electron beams carrying orbital angular momentum *Nature* **464** 737-9
[196] Verbeeck J, Tian H and Schattschneider P 2010 Production and application of electron vortex beams *Nature* **467** 301-4
[197] Schattschneider P, Schachinger T, Stoger-Pollach M, Loffler S, Steiger-Thirsfeld A, Bliokh K Y and Nori F 2014 Imaging the dynamics of free-electron Landau states *Nature Communications* **5** 4586
[198] McMorran B J, Agrawal A, Anderson I M, Herzing A A, Lezec H J, McClelland J J and Unguris J 2011 Electron vortex beams with high quanta of orbital angular momentum *Science* **331** 192-5
[199] Grillo V, Gazzadi G C, Mafakheri E, Frabboni S, Karimi E and Boyd R W 2015 Holographic generation of highly twisted electron beams *Phys. Rev. Lett.* **114** 034801
[200] Verbeeck J, Tian H and Van Tendeloo G 2013 How to Manipulate nanoparticles with an electron beam? *Adv. Mater.* **25** 1114-7
[201] Schattschneider P, Ennen I, Loffler S, Stoger-Pollach M and Verbeeck J 2010 Circular dichroism in the electron microscope: Progress and applications (invited) *J. Appl. Phys.* **107** 09D301
[202] Beche A, Van Boxem R, Van Tendeloo G and Verbeeck J 2014 Magnetic monopole field exposed by electrons *Nature Physics* **10** 26-9
[203] Esarey E, Schroeder C B and Leemans W P 2009 Physics of laser-driven plasma-based electron accelerators *Rev. Mod. Phys.* **81** 1229-85

[204] Emma P, Huang Z, Kim K J and Piot P 2006 Transverse-to-longitudinal emittance exchange to improve performance of high-gain free-electron lasers *Phys. Rev. ST Accel. Beam* **9** 100702

[205] Gover A 2005 Superradiant and stimulated-superradiant emission in prebunched electron-beam radiators. I. Formulation *Phys. Rev. ST Accel. Beam* **8** 030701

[206] Gover A, Dyunin E, Lurie Y, Pinhasi Y and Krongauz M V 2005 Superradiant and stimulated-superradiant emission in prebunched electron-beam radiators. II. Radiation enhancement schemes *Phys. Rev. ST Accel. Beam* **8** 030702

[207] Liu Y, Bogacz S A, Cline D B and Wang X J 1997 Micro-bunching diagnostics for inverse Cherenkov acceleration by coherent transition radiation *Nucl. Instrum. Methods Phys. Res., Sect. A* **386** 295-300

[208] Sharma S C and Malik P 2015 The effect of beam pre-bunching on the excitation of terahertz plasmons in a parallel plane guiding system *Phys. Plasmas* **22** 043301

[209] Eichenbaum A L 1999 Traveling wave prebunching of electron beams for free electron masers *IEEE Trans. Plasma Sci.* **27** 568-74

[210] Arbel M, Eichenbaum A L, Pinhasi Y, Lurie Y, Tecimer M, Abramovich A, Kleinman H, Yakover I M and Gover A 2000 Super-radiance in a prebunched beam free electron maser *Nucl. Instrum. Methods Phys. Res., Sect. A* **445** 247-52

[211] Smirnova O, Spanner M and Ivanov M 2008 Analytical solutions for strong field-driven atomic and molecular one- and two-electron continua and applications to strong-field problems *Phys. Rev. A* **77** 033407

[212] Wolkow D M 1935 Über eine Klasse von Lösungen der Diracschen Gleichung *Zeitschrift für Physik* **94** 250-60

[213] Madsen L B 2005 Strong-field approximation in laser-assisted dynamics *Am. J. Phys.* **73** 57-62

[214] Ohtsuki Y H 1970 Theory of Low Energy Electron Diffraction .3. Inelastic Scattering *J. Phys. Soc. Jpn.* **29** 398

[215] Hofmann F and Smith H P 1967 Dynamical Theory of Low-Energy Electron Diffraction *Phys. Rev. Lett.* **19** 1472

[216] White A J, Sukharev M and Galperin M 2012 Molecular nanoplasmonics: Self-consistent electrodynamics in current-carrying junctions *Phys. Rev. B* **86** 205324

[217] Sukharev M and Nitzan A 2011 Numerical studies of the interaction of an atomic sample with the electromagnetic field in two dimensions *Phys. Rev. A* **84** 043802

[218] Mencarelli D, Rozzi T, Maccari L, Di Donato A and Farina M 2007 Electronic properties of carbon nanotubes investigated by means of standard electromagnetic simulators *Phys. Rev. B* **75** 085402

[219] Verboncoeur J P 2005 Particle simulation of plasmas: review and advances *Plasma Phys. Control. Fusion* **47** A231

[220] Zhou J, Hu M, Zhang Y X, Zhang P, Liu W H and Liu S G 2011 Numerical analysis of electron-induced surface plasmon excitation using the FDTD method *J. Opt.* **13** 035003

[221] Buneman O 1959 Dissipation of currents in ionized media *Phys. Rev.* **115** 503-17

[222] Dawson J 1962 One-dimensional plasma model *Phys. Fluids* **5** 445-59

[223] Lee R, Boris J P and Haber I 1972 Electromagnetic simulation codes for relativistic plasmas *Am. Phys. Soc.* **17** 1048

[224] Vay J L 2008 Simulation of beams or plasmas crossing at relativistic velocity *Phys. Plasmas* **15**

[225] http://www.feynmanlectures.caltech.edu/II_28.html.

[226] John D L, Castro L C, Pereira P J S and Pulfrey D L 2004 A Schrodinger-Poisson solver for modeling carbon nanotube FETs *Nsti Nanotech 2004, Vol 3, Technical Proceedings* 65-8

[227] Russer P and Siart U 2008 *Time domain methods in electrodynamics: A tribute to Wolfgang J. R. Hoefer* (Berlin, Heidelberg: Springer-Verlag)

[228]    Alu A, Silveirinha M G, Salandrino A and Engheta N 2007 Epsilon-near-zero metamaterials and electromagnetic sources: Tailoring the radiation phase pattern *Phys. Rev. B* **75** 155410